\documentclass[11pt]{article}

\usepackage{graphicx,rotating,slashed,amsmath,xcolor,amssymb,amsfonts,colortbl,cite,setspace}
\usepackage{color}	 
\usepackage{jcappub}
\usepackage[normalem]{ulem}
\usepackage{hyperref}
\pdfoutput=1
\makeatletter 
\usepackage{bm}        % for math

\hypersetup{colorlinks,bookmarksopen,bookmarksnumbered,
linkcolor=blus,pdfstartview=FitH,urlcolor=rossos,citecolor=verde}
\allowdisplaybreaks

\def\lsim{\mathrel{\rlap{\lower3pt\hbox{\hskip0pt$\sim$}}
   \raise1pt\hbox{$<$}}}         %less than or approx. symbol
\def\gsim{\mathrel{\rlap{\lower4pt\hbox{\hskip1pt$\sim$}}
   \raise1pt\hbox{$>$}}}         %greater than or approx. symbol

 \newcommand{\sfootnote}[1]{}
\definecolor{bluc}{cmyk}{1,1,0,0.1}
\definecolor{rossoCP3}{cmyk}{0,.88,.77,.40}
\definecolor{rosso}{cmyk}{0,1,1,0.4}
\definecolor{rossos}{cmyk}{0,1,1,0.55}
\definecolor{rossoc}{cmyk}{0,1,1,0.2}
\definecolor{verdes}{cmyk}{0.92,0,0.59,0.4}

\def\noi{{\noindent}}

\def\blu{\color{blue}}

\hypersetup{colorlinks, bookmarksopen, bookmarksnumbered,
citecolor=verdes, linkcolor=bluc, pdfstartview=FitH, urlcolor=rossos}

\newcommand{\mio}[1]{}

\definecolor{Gray}{gray}{0.95}

\usepackage[makeroom]{cancel}
\makeatletter
\def\cantox@vector#1#2#3#4#5#6#7#8{%
  \dimen@.5\p@
  \setbox\z@\vbox{\boxmaxdepth.5\p@
   \hbox{\kern-1.2\p@\kern#1\dimen@$#7{#8}\m@th$}}%
  \ifx\canto@fil\hidewidth  \wd\z@\z@ \else \kern-#6\unitlength \fi
  \ooalign{%
    \canto@fil$\m@th \CancelColor
    \vcenter{\hbox{\dimen@#6\unitlength \kern\dimen@
      \multiply\dimen@#4\divide\dimen@#3 \vrule\@depth\dimen@\@width\z@
      \vector(#3,-#4){#5}%
    }}_{\raise-#2\dimen@\copy\z@\kern-\scriptspace}$%
    \canto@fil \cr
    \hfil \box\@tempboxa \kern\wd\z@ \hfil \cr}}
\def\bcancelto#1#2{\let\canto@vector\cantox@vector\cancelto{#1}{#2}}
\makeatother

\usepackage{soul}
\usepackage{hyperref}
\usepackage{multicol}
\usepackage{dsfont}
\usepackage{color}
\definecolor{rosso}{cmyk}{0,1,1,0.4}
\definecolor{rossos}{cmyk}{0,1,1,0.55}
\definecolor{rossoc}{cmyk}{0,1,1,0.2}
\definecolor{blu}{cmyk}{1,1,0,0.3}
\definecolor{blus}{cmyk}{1,1,0,0.6}
\definecolor{bluc}{cmyk}{1,1,0,0.1}
\definecolor{verde}{cmyk}{0.92,0,0.59,0.25}
\definecolor{verdec}{cmyk}{0.92,0,0.59,0.15}
\definecolor{verdes}{cmyk}{0.92,0,0.59,0.4}

\renewcommand\&{&}

\def\circa#1{\,\raise.3ex\hbox{$#1$\kern-.75em\lower1ex\hbox{$\sim$}}\,}

\newcommand{\beq}{\begin{equation}}
\newcommand{\eeq}{\end{equation}}

\newcommand{\bea}{\begin{eqnarray}}
\newcommand{\eea}{\end{eqnarray}}
\newcommand{\be}{\begin{equation}}
\newcommand{\ee}{\end{equation}}
\newfam\rsfsfam
\def\mathscr#1{{\fam\rsfsfam\relax#1}}

\def\circa#1{\,\raise.3ex\hbox{$#1$\kern-.75em\lower1ex\hbox{$\sim$}}\,}
\makeatletter

\def\hhref#1{\href{http://arxiv.org/abs/#1}{arXiv:#1}} % in bibliography

\newcommand{\doi}[1]{\href{http://dx.doi.org/#1}{[doi]}}

\def\hhref#1{\href{http://arxiv.org/abs/#1}{arXiv:#1}} 
 
\def\art{\@ifnextchar[{\eart}{\oart}}
\def\eart[#1]#2#3#4#5#6{{\rm #2}, {\em #3 \bf #4} {\rm (#6) #5} ({\em #1})}

\def\article{\@ifnextchar[{\earticle}{\oarticle}}
\def\oarticle#1#2#3#4#5#6{{\rm #1}, {\em ``#6''}, {\rm #2 #3 (#5) #4}}
\def\earticle[#1]#2#3#4#5#6#7{{\rm #2}, {\em ``#7''}, {\rm #3 #4 (#6) #5}  [\hhref{#1}]}
\def\hepart[#1]#2{{\rm #2, \em#1}}
\def\heparticle[#1]#2#3{#2, {\em ``#3''} [\hhref{#1}]}

\newcounter{alphaequation}[equation]
\def\thealphaequation{\theequation\hbox to
0.6em{\hfil\alph{alphaequation}\hfil}}
\def\eqnsystem#1{
\def\@eqnnum{{\rm (\thealphaequation)}}
\def\@@eqncr{\let\@tempa\relax \ifcase\@eqcnt \def\@tempa{& & &} \or
  \def\@tempa{& &}\or \def\@tempa{&}\fi\@tempa
  \if@eqnsw\@eqnnum\refstepcounter{alphaequation}\fi
\global\@eqnswtrue\global\@eqcnt=0\cr}
\refstepcounter{equation} \let\@currentlabel\theequation \def\@tempb{#1}
\ifx\@tempb\empty\else\label{#1}\fi
\refstepcounter{alphaequation}
\let\@currentlabel\thealphaequation
\global\@eqnswtrue\global\@eqcnt=0 \tabskip\@centering\let\\=\@eqncr
$$\halign to \displaywidth\bgroup \@eqnsel\hskip\@centering
$\displaystyle\tabskip\z@{##}$&\global\@eqcnt\@ne
\hskip2\arraycolsep\hfil${##}$\hfil& \global\@eqcnt\tw@\hskip2\arraycolsep
$\displaystyle\tabskip\z@{##}$\hfil
\tabskip\@centering&\llap{##}\tabskip\z@\cr}
\def\endeqnsystem{\@@eqncr\egroup$$\global\@ignoretrue} \makeatother

\definecolor{fiorentina}{rgb}{.5,0,.5}

\begin{document}

\centerline{\title{
%{\blu Quasi single tensor field inflation\\}
Probing the inflationary particle content: extra spin-2 field}} 

\author{\hspace{-0.7cm}Emanuela Dimastrogiovanni$\,^{a,b}$, Matteo Fasiello$\,^{c,d}$, Gianmassimo Tasinato$\,^{e}$}
\affiliation{\hspace{-0.7cm}$^{a}$ CERCA and Department of Physics, Case Western Reserve University,
Cleveland, OH, 44106, U.S.A. }
\affiliation{\hspace{-0.7cm}$^{b}$ Perimeter Institute for Theoretical Physics,
31 Caroline Street North, Waterloo, N2L 2Y5, Canada.}
\affiliation{\hspace{-0.7cm}$^{c}$
Institute of Cosmology and Gravitation, University of Portsmouth, Portsmouth, PO1 3FX, UK.}
\affiliation{\hspace{-0.7cm}$^{d}$ Department of Physics, Stanford University, Stanford, CA, 94306, U.S.A.}
\affiliation{\hspace{-0.7cm}$^{e}$ Department of Physics, Swansea University, Swansea, SA2 8PP, United Kingdom}

%\date{\today}

%%%%%%%%%%%%%%%%%%%%%%%%%%%%%%%%%%%%%%%%%%%%%%%%%%%%%%%%%%%%%%%%%%%%%
%%%% Abstract

\hspace{-0.7cm}\abstract{We study how inflationary observables associated with primordial tensor modes are  affected by coupling the minimal field content with an extra spin-2 particle during inflation. We work with a model that is ghost-free at the fully non-linear level and show how the new degrees of freedom modify standard consistency relations for the tensor bispectrum. The extra interacting spin-2 field is necessarily massive and unitarity dictates its mass be in the $m_{\rm}\gtrsim H$ range. Despite the fact that this bound  selects a decaying solution for the corresponding tensor mode, cosmological correlators still carry the imprints of such \textit{fossil} fields. Remarkably, fossil(s) of spin $\geq 1$
generate distinctive anisotropies in observables such as the tensor power spectrum. We show how this plays out in our set-up.

 }

\maketitle

\section{Introduction}

Cosmological inflation predicts the existence of a  yet-unobserved stochastic background of primordial tensor modes,  whose detection   represents one of the main challenges for future observational cosmology  (see e.g. \cite{Kamionkowski:2015yta,Guzzetti:2016mkm,Caprini:2018mtu} for recent reviews). According to the simplest realizations of inflation, the primordial gravitational waves background is nearly scale invariant and almost Gaussian, with CMB polarization measurements being potentially one of its most sensitive probe. Nevertheless, in several  scenarios, well-motivated by particle physics, there arise the possibility that primordial gravitational waves (GW) can be detected at interferometer scales \cite{Bartolo:2016ami}. Given that  primordial GW represent an additional window on the dynamics of inflation, and can offer important
clues on its ultraviolet completion, it is essential to   explore  all possibilities in view of a future detection \cite{Arkani-Hamed:2015bza}. \\ 
\indent Besides the GW signal due to standard vacuum fluctuations, there exist several different mechanisms that  can source primordial tensor waves. One possibility stems from specific couplings (typically non-minimal) of the inflaton with additional fields  (from scalars to vectors, all the way to higher spin fields, see e.g. \cite{Cook:2013xea}). Such models often lead to (controlled) instabilities and particle production affecting the tensor sector both at CMB and interferometer scales. Alternatively, the primordial tensor statistics can be modified by breaking space-time symmetries during inflation,  either spontaneously as in (super)-solid inflation \cite{Endlich:2012pz}, or directly as in Ho\v rava-Lifshitz-motivated inflationary systems (see e.g. \cite{Takahashi:2009wc}). \\
\indent In addition to the power spectrum, its spectral index, and possible chiral properties, an excellent probe of extra dynamics during inflation is the squeezed (soft) configuration of the bispectrum. Indeed this observable is sensitive to the \textit{mass}, the \textit{spin} and the \textit{coupling} of any additional particle content. For example, the effect of an extra light scalar on the predictions of minimal single-field slow-roll (SFSR) inflation is to enhance the bispectrum in the squeezed limit. From this property stems the well-known identification of a $f^{\rm loc}_{\rm nl}\gtrsim 1$ detection as a ``smoking gun" for multi-field inflation \cite{Creminelli:2004yq}. Crucially, the information potentially stored  in the bispectrum spectrum extends much further than that. Indeed, as one considers extra particle content of increasing spin, the corresponding bispectrum acquires a non-trivial angular dependence. At the same time, the bispectrum scaling with respect to soft (hard) momenta reveals information on the extra particle(s) mass. This latter property is rather remarkable: even if the field content beyond SFSR is massive to the level $m_{\rm extra}\sim H$, and therefore short-lived, its imprints are \textit{fossilised} \cite{Jeong:2012df} in the bispectrum and remain accessible today. This is especially important in view of the fact that the existence of irreducible unitary representation of the spacetime isometries group strongly constraints the parameter space of theories with extra spin$\geq$2 fields.\\
\indent In this work  we extend the minimal inflationary scenario by coupling it with an extra spin-2 field \cite{deRham:2010ik,deRham:2010kj,Hassan:2011zd} and focus on the consequences for the tensor sector.  This particle content, comprising an additional tensor with a mass range of order the Hubble parameter, is reminiscent of a well-studied  scenario known as quasi-single field inflation (QsF) \cite{Chen:2009we,Chen:2009zp}, where the additional field is a massive $m_{\rm\sigma}\sim H$ scalar. In this sense, our set-up can be thought of as the tensor sector counterpart of QsF. We refer the reader to the literature \cite{Kehagias:2017cym,Bartolo:2017sbu} (see also \cite{Lee:2016vti}) for related studies of (extra) higher spin fields populating inflation. In choosing the extra field to be a spin-2 particle, one can rely on an extensive body of work \cite{deRham:2014zqa} (see in particular \cite{Lagos:2014lca,Johnson:2015tfa,Cusin:2015pya,Biagetti:2017viz} for use in the inflationary context) spanning also models of the late-time universe. As we shall see, any (interacting) addition to the tensor sector of general relativity is necessarily massive. We plan to make use of a fully-non-linear ghost-free couplings between the tensor sectors known as \textit{dRGT} interactions \cite{deRham:2010kj}. In doing so, we will exploit some of the model most desirable properties:  

 \begin{itemize}
 \item The structure of the theory is engineered to guarantee the absence of Ostrogradsky instabilities. In fact, the model of \cite{Hassan:2011zd} is the only known example of  multi-metric theory consistent at the fully non-linear level,  and precisely dictates the coupling between massive and massless tensor fluctuations\footnote{Strictly speaking, the tensor mass eigenstates in the theory are time-dependent.}.

 \item The strong coupling scale (and therefore also the UV cut-off) of the theory is naturally  much larger (see \textit{Section} \ref{scs}) than our working energy, set by the Hubble parameter $H$. This, together with the fact that the model is automatically ghost-free at fully non-linear level, implies that one can explore a regime in parameter space that cannot be otherwise probed using a fully general  EFT approach\footnote{In other words, there exists a subset of operators/non-linear interactions, w.r.t. those that would appear in a broader EFT treatment, that can be as important as the linear theory at a specific scale $\Lambda$ without necessarily exciting any ghostly degree of freedom.}.

 \item Adding a massive spin-2 to the minimal inflationary scenario comes with an extra five degrees of freedom: two tensors, two vectors and a scalar. When the non-linearities in the dRGT interactions become important, the effective coupling to matter (including the inflaton) of the extra vectors/scalar is strongly suppressed: this is what is known as the Vainshtein screening mechanism at work \cite{Babichev:2013usa}. On the other hand, screening does not affect the extra tensor sector, precisely  the one we shall focus on in this work. The key observable will be the tensor three-point function, the smallest $n$-point function to directly probe non-Gaussianities in the theory and imprints of extra modes therein. We will study the contributions to the bispectrum due to the new \textit{non-derivative} dRGT interactions and show how these modify   the standard tensor consistency condition  in play for SFSR inflation. 
\end{itemize}
    
\noindent Non-derivative interactions tend to enhance the so-called \textit{local} contributions to the bispectrum or lead to intermediate (i.e. interpolating between different configurations) shapes for non-Gaussianities. This is well-studied for the scalar sector (see e.g. the review in  \cite{Chen:2010xka}), and is also expected to hold for the tensor sector.  Our results provide a   theoretical and phenomenological characterization of tensor non-Gaussianities beyond SFSR that is essential in view of future experimental probes \cite{Meerburg:2016ecv}.\\

\noi This paper is organized as follows: in \textit{Section} \ref{hellohello} we introduce our non-minimal inflationary scenario and study the dynamics of tensor fluctuations around de Sitter space (a proxy for inflation) up to fourth order in perturbations. We examine the bounds on fields mass range that stem from requiring this to be a consistent set-up; in \textit{Section}  \ref{sec-obs} we analyse how the presence of new tensor modes affects cosmological correlators, and the consequences for tensor non-Gaussianity;  \textit{Section}  \ref{tcr} is devoted to more general considerations on tensor consistency relations in  light of our findings, and on observational prospects for the bispectrum resulting from modifications/violations of consistency relations; we conclude in \textit{Section} \ref{sec-con}, whilst technical details of the calculations can be found in the \textit{Appendices}.

\section{The model}
\label{hellohello}
Our set-up consists of (i) the standard action \footnote{Although for simplicity we discuss a
minimal  single field inflationary scenario, most of the arguments we develop here also apply
to  richer systems including additional scalar (and vector) fields. 
} for the inflaton field in single field slow-roll inflation (SFSR) 
\bea
\label{minimal}
S_{\phi}=\int d^4x \sqrt{-g} \left[ - \frac12 (\partial_{\mu}\phi)^2 - V(\phi) \right]\,,
\eea
where the inflaton $\phi$ (and, in general, any matter content) is minimally coupled to gravity, described by the metric $g_{\mu\nu}$.
 We further consider (ii) an additional spin-2 field, making use of the fact that the only known system consistently coupling two spin-2 tensor modes, here called
$g_{\mu\nu}$ and $f_{\mu\nu}$, is described by the action of  \cite{Hassan:2011zd}:

\begin{equation}
\label{inact1}
S=\int d^4x \left[\sqrt{-g}\,M_{p}^2\,R[g]+\sqrt{-f}\,M_{f}^2\,R[f]-\sqrt{-g}\,m^2 { M^2} \sum_{n=0}^{4}\beta_{n}\,{\cal E}_{n}\left(\sqrt{g^{-1}f} \right)\right]\,,
\end{equation}
 where the combinations 
 ${\cal E}_{n}(\mathbb{X})$ (with $\mathbb{X}$ an arbitrary matrix) are defined
as
\bea
&&\mathcal{E}_0(\mathbb{X})=1, \quad \mathcal{E}_1(\mathbb{X})={\rm Tr}(\mathbb{X}) \equiv [\mathbb{X}], \quad \mathcal{E}_2(\mathbb{X})=\frac{1}{2}\left( [\mathbb{X}]^2 - [\mathbb{X}^2]  \right),\nonumber\\
&&\mathcal{E}_3(\mathbb{X})=\frac{1}{3!}\left( [\mathbb{X}]^3 -3 [\mathbb{X}^2][\mathbb{X}]+2 [\mathbb{X}^3]  \right), \nonumber\\
&&\mathcal{E}_4(\mathbb{X})=\frac{1}{4!}\left( [\mathbb{X}]^4 -6 [\mathbb{X}^2][\mathbb{X}]^2 +8 [\mathbb{X}^3][\mathbb{X}] +3 [\mathbb{X}^2]^2-6[\mathbb{X}^4]   \right)\, .
\label{poly}
\eea
The action  \eqref{inact1} contains the Einstein-Hilbert  terms for the metrics $f$ and $g$ (with
$M_f$ and $M_p$ denoting the corresponding
Planck masses), a potential term controlled by the parameter $m$ parameterizing the mass in the tensor sector, and
five dimensionless constant parameters $\beta_i$. For
convenience -- as well as to keep the $f \leftrightarrow g$  identification\footnote{Before coupling to matter, the action is invariant under the simultaneous exchange $M_p\leftrightarrow M_f\,;\, g\leftrightarrow f\,;\beta_n\leftrightarrow\beta_{4-n} $. } of \cite{Hassan:2011zd} manifest --  we use the parameter $M$ defined as
\be \label{defM}
M^2=\frac{M_p^2 M_f^2}{ M_p^2 +M_f^2}\,.
\ee
As a proxy for inflation, we consider de Sitter solutions for both background metrics $f,g$. The corresponding background equations of motion ($eom$) are
\begin{eqnarray}
&&3\, H^2 M_{p}^{2}=m^2 { M^2}\left(\frac{1}{2}\beta_{0}+\frac{3}{2}\beta_{1}+\frac{3}{2}\beta_{2}+\frac{1}{2}\beta_{3}\right)\nonumber\\
&&3\, H^2 M_{f}^{2}=m^2 M^2\left(\frac{1}{2}\beta_{1}+\frac{3}{2}\beta_{2}+\frac{3}{2}\beta_{3}+\frac{1}{2}\beta_{4}\right)\,.
\label{baeqs}
\end{eqnarray}

\noi Note that in Eq.(\ref{baeqs}) is implicit the choice of the healthy branch \cite{deRham:2014zqa} of solutions ($H_g/H_f=b/a$) and we have in particular considered the case $H_g=H_f=H$. We write the metric fluctuations as
\bea
g_{\mu\nu} \,d x^\mu d x^\nu &=&-d t^2+a^2(t) \left[ \delta_{ij}+g_{ij}(t,\vec x)\right]\,d x^i d x^j
\\
f_{\mu\nu} \,d x^\mu d x^\nu&=&
-d t^2+a^2(t) \left[ \delta_{ij}+f_{ij}(t,\vec x)\right]\,d x^i d x^j
\eea
 As anticipated, we will focus our analysis on the tensor sector and in particular on the traceless transverse components of the metrics $f,g$.  Although we shall not consider them directly, we nevertheless ought to ensure that we explore a region in parameter space for the theory where unitarity is preserved also in the helicity-0 and helicity-1 sector. To this aim, we will enforce  on the parameters $m$ and $\beta_i$ the \textit{Higuchi} bound of Section \ref{sec-hig}. In order to extract the observables of interest here, namely  the tensor power spectrum and bispectrum up to one-loop order, it will be necessary to expand the action up to fourth order in perturbations. The new physics w.r.t. SFSR is in the non-derivative interactions making up the action that we schematically write as:
 \be
 S_{int}\,=\,\int d t\,d^3 x\,a^{3}(t)\,\left[ 
 {\cal L}_2+ {\cal L}_3+ {\cal L}_4 
 \right]\; .
 \ee
The  Lagrangian densities at each order read

\bea\label{lagl2}
 {\cal L}_2&=& -
\frac{m^2\,M_{}^2}{8}\,\left( \beta_1+2 \beta_2+\beta_3\right)
\, {\rm Tr} \left[ (f-g)^2 \right]\; ,
\\
{\cal L}_3&=& \frac{m^2\,M_{}^2}{48}
\,\Big[
\left( 7 \beta_1 +12 \beta_2 +5\beta_3\right) {\rm Tr} \left[g^3\right]
-3 \left(  \beta_1 +4 \beta_2 +3\beta_3\right)
 {\rm Tr} \left[ g f^2 \right]
 \nonumber
\\
&&\hskip2.5cm-3
\left( 3 \beta_1+4 \beta_2 +\beta_3\right) {\rm Tr} \left[ g^2 f \right]
+\left( 5 \beta_1+12 \beta_2 +7 \beta_3\right) {\rm Tr} \left[  f^3 \right]
\Big]\; ,
\\
{\cal L}_4&=&- \frac{m^2\,M_{}^2}{128}\,
\left( \beta_1+2 \beta_2+\beta_3\right)
\,\Big[
3 \,{\rm Tr} \left[g^4\right]-4 \,
 {\rm Tr} \left[ g^3 f \right]
 \nonumber
\\
&&\hskip2.5cm+
18 \, {\rm Tr} \left[ g^2 f^2 \right]
-4  \, {\rm Tr} \left[ g f^3 \right]
 \nonumber
\\
&&\hskip2.5cm
+ 3 \, \,{\rm Tr} \left[  f^4 \right]
-8 \,\left( {\rm Tr} \left[  f g \right] \right)^2
\Big]\; ,
\label{lagl4}
\eea

\noi 
where ${\rm Tr} \left[  f \right]\,=\, {\rm Tr} \left[  f_{\mu\nu} \right]$ etc. Notice that the new interactions described by the potential term in Eq.~\eqref{inact1} and in Eqs. \eqref{lagl2}-\eqref{lagl4} do not contain any derivatives. Indeed, there cannot be derivative couplings among spin-2 fields that are  consistent at the fully non-linear level \cite{deRham:2013tfa}. Such couplings are of course still allowed in the purely EFT sense of e.g. \cite{Lee:2016vti}. Despite the compact expression for each $\mathcal{L}_n$, the $(f,g)$ basis mixes the two metrics already at second order, as in
\be
M_{p}^2\,{\cal L}_{kin}(g)+M_{f}^2\,{\cal L}_{kin}(f)-\frac{1}{8}m^2 M^2\left(\beta_{1}+2\,\beta_{2}+\beta_{3}\right)\left(g^{}_{ij}-f_{ij}\right)^2\, ,
\ee
with ${\cal L}_{kin}$ indicating the standard quadratic kinetic terms for the tensors. It is convenient to use an alternative basis that decouples the fields in $\mathcal{L}_{2}$. In order to do so, one may first canonically normalize the spin-2 kinetic  terms, rescaling  fields according to
\bea
g_{ij}\to \tilde g_{ij}\equiv g_{ij} M_p\;\,
\\
f_{ij}\to \tilde f_{ij} \equiv f_{ij} M_f\; ,
\eea
and then considering a rotation in field space so as to diagonalize the potential: 
\bea
\tilde g_{ij}&=&\cos \alpha  \,  \sigma_{ij} +\sin \alpha \,  \gamma_{ij}\;\,
\\
\tilde f_{ij}&=&-\sin \alpha  \, \sigma_{ij} +\cos  \alpha \, \gamma_{ij}\;,
\eea
with
\be
\alpha\,=\,\arccos \left[\frac{M_f}{\sqrt{M_p^2+M_f^2}} \right]\; .
\ee
The quadratic Lagrangian in the new basis reads
\be
{\cal L}_{kin}( \sigma)+{\cal L}_{kin}( \gamma )-\frac{1}{8}m^2  \,{\tilde{\beta}}\,
 \sigma^2\; ,
\ee
with  $\tilde{\beta}$ given by the combination:
\be \label{deftb}
{\tilde{\beta}}\equiv\left(\beta_{1}+2\,\beta_{2}+\beta_{3}\right)\; .
\ee
Interactions between the massless field $\gamma_{ij}$ and
the massive $\sigma_{ij}$ particle start now at third order, and are regulated by the following cubic Lagrangian 
\bea
\mathcal{L}_{3}=&&\frac{m^2  M {\tilde{\beta}} }{4 M_f M_p}  \sigma ^2  \gamma + \frac{m^2 M  }{48\,M_f^2 M_p^2}\Big[ (7\beta_1+12\beta_2+5\beta_3)M_f^2-(5\beta_1+12\beta_2+7\beta_3)M_p^2 \Big]   \sigma^3 \;.\;
\label{cubicL}
\eea
The corresponding expression for the quartic non-derivative interactions is more lengthy  and can be found in Eq.(\ref{qL})  of Appendix \ref{app-qua}. It is sufficient at this stage to point out that all its terms are proportional to a unique combination of the parameters $\beta_i$ corresponding to $\tilde\beta$ as per Eq.~\eqref{deftb}.

\subsection{The structure of the interactions}
\label{sec-hig}

At quadratic level we identified a basis where massless and massive tensor modes propagate independently, the mass being proportional a linear  combination of the $\beta_n$ coefficients we call ${\tilde{\beta}}$, see Eq.~\eqref{deftb}. At cubic and quartic order interactions among the two fields $\gamma,\sigma$ are controlled by ${\tilde{\beta}}$ and there are no new self-interactions for the massless mode $\gamma$ besides the ones already present in GR. These features can be understood quite generally within the context of the theorem in \cite{Boulanger:2000rq}: it states that, in the limit in which the massive field $\sigma$ becomes massless (that is, ${\tilde{\beta}}\to0$),  interactions among the two massless
fields must vanish. As we shall see, an implication of this result for our set-up is that the new contributions to the power spectrum and bispectrum appear starting with loop diagrams, {and, at leading order,} depend on the dimensionless parameter ${\tilde{\beta}}$.

\subsubsection*{Bounds on the parameters space}

We provide  here a brief discussion on the bounds that limit our analysis to a specific region of the parameter space of the theory. First, we recall \cite{Wigner:1939cj} that the notion of particles is best defined by unitary irreducible representation of (in our case) the de Sitter isometry group. These identify, for massive\footnote{ There is a discrete set of allowed values for the mass to which corresponds, at least linearly, an enhanced symmetry in the theory and therefore less propagating degrees of freedom (four). This  is known as a \textit{partially massless} system \cite{Deser:2001wx}. However, a consistent non-linear theory of partially massless gravity, if it exists,  has yet to be found.} spin-2 fields, the condition \cite{Fasiello:2012rw,Fasiello:2013woa}:  
\bea
\frac{m^2 M^2}{H^2 M_p^2} \tilde{\beta} \left(1+\frac{M_p^2}{M_f^2}\right)
\,=\,\frac{m^2\,\tilde \beta}{H^2}
\geq4 \Bigg|_{\rm Higuchi}\; ,
\label{higuchi1}
\eea 
where have used the definition in Eq.~\eqref{defM} for $M$. This condition
 reduces to the well-know Higuchi \cite{Higuchi:1986py} bound  $m^2\geq 2H^2$ in the linear massive gravity limit ($
 M_f\rightarrow \infty 
 \; ;
 \tilde{\beta} \rightarrow 1/2$). 
We note in passing that the unitarity bound on the mass of the (extra) tensor mode confines us to the same $m\gtrsim H$ playground that defines the parameter space of the extra scalar mode in QsF inflation. \\
\indent The origin of the next bound is rather different, and the corresponding inequality may in principle be relaxed  depending on the set-up under consideration. In obtaining the solution for the $\sigma$ wavefunction in Eq.~(\ref{solgm}), we  assume  a real-valued $\tilde{\nu}$.  Were we to allow, as it is certainly possible\footnote{See for example \cite{Lee:2016vti}, according to whose nomenclature our analysis is on the mass values (those corresponding to $\tilde{\nu}\in {\rm Reals}$) belonging to the complementary series, and an imaginary  $\tilde{\nu}$ corresponds to the principal series.}, an imaginary-valued $\tilde{\nu}$, one  typically  expects a Boltzmann $e^{-m/H}$ suppression in correlation functions due to the presence of very massive modes (see \cite{Chen:2009zp} for a detailed discussion on this point). Written along the lines of Eq. (\ref{higuchi1}), a real-valued $\tilde{\nu}$ corresponds to
 \bea
\quad \frac{m^2 M^2}{H^2 M_p^2}{\tilde{\beta}} \left(1+\frac{M_p^2}{M_f^2}\right)
\,=\,\frac{m^2\,\tilde \beta}{H^2}\,
\leq\, \frac{9}{2} \Bigg|_{\rm EOM}\; .
\label{nutilde}
\eea

\noi In what follows we will be assuming both inequalities \eqref{higuchi1}  and  \eqref{nutilde}, and consider a mass range

\be \label{finin}
 4\le \frac{m^2\,\tilde \beta}{H^2} \le \frac{9}{2}\;,
\ee
although, as mentioned above, the upper bound is not  strictly necessary and may be relaxed.

\subsubsection*{Strong coupling scale}
\label{scs}
In order to expound on the strong coupling scale of the model under scrutiny, we need to make contact with the non-linear massive gravity theory of \cite{deRham:2010kj}. In such context, the so-called \textit{naive} strong coupling scale is $\Lambda_3\equiv  (m^2 M_p)^{1/3}$. We note that it emerges most clearly in a specific scaling limit of the theory known as decoupling limit (DL). The DL analysis shows how the smallest strong coupling scale has to do with the helicity-0 mode interactions. Indeed, for purely tensor modes one would expect that scale to be $M_p$, just as is the case for general relativity.\\
\indent On non-trivial backgrounds the non-linearities in the massive theory actually re-dress the effective strong coupling scale, always in the direction of making it larger, $\Lambda_{*}\sim \sqrt{Z}\,\Lambda_3$, where $Z$ measures the non-linear terms contributions to the standard kinetic term for the helicity-0 mode. In Eq.~(\ref{inact1}) both metrics $g,f$ are dynamical and, as a result, the naive strong coupling scale becomes $\Lambda={\rm Min}\{\Lambda_g, \Lambda_f\}$ with $\Lambda_{g,f}= (m^2 M_{g,f})^{1/3}$ . Given the bounds on $m$ we discuss in the previous section and the fact that we assume $M_{g,f}\gg H$, one may conclude we are safely away even from what is only the naive strong coupling scale in our set-up. It is worth pointing out that it would be hard to put together a scale $H$ compatible with inflation, such as the one we require here, with the use of a massive graviton as a mechanism for late-time acceleration. We make no such attempt here, and view our  Eqs.~(\ref{minimal})-(\ref{inact1}) as simply describing the (healthy) field content of a minimal inflation model enriched by a spin-2 particle.

%%%%%%%%%%%%%
\section{Observables in the tensor sector}
\label{sec-obs}
%%%%%%%%%%%%%%

We first quantize tensor fluctuations around de Sitter space and then investigate the properties of the  power spectrum and bispectrum. Employing the $(\gamma,\sigma$) basis  the massless and massive tensor modes are decomposed in Fourier space as
\bea
\hat{h}_{ij}(\textbf{x},\tau)&=&\int\frac{d^3k}{(2\pi)^3} e^{i\textbf{k}\cdot\textbf{x}}\sum_{\lambda}\epsilon^{\lambda}_{ij}(\hat{k})\hat{h}^{\lambda}(\textbf{k},\tau)\,,
\\
\hat{\sigma}_{ij}(\textbf{x},\tau)&=&\int\frac{d^3k}{(2\pi)^3} e^{i\textbf{k}\cdot\textbf{x}}\sum_{\lambda}\epsilon^{\lambda}_{ij}(\hat{k})\hat{\sigma}^{\lambda}(\textbf{k},\tau)\,,
\eea
where the polarization tensors defined in Appendix~\ref{appB}. The operators $\hat h^\lambda$, $\hat \sigma^{\lambda}$ are written in terms of standard creation/annihilation operators, satisfying the standard commutation relations, and the mode functions as in
\bea
\hat{\gamma}^{\lambda}(\textbf{k},\tau)=a_{\textbf{k}}^{\lambda}\, \gamma(k,\tau)+a_{-\textbf{k}}^{\lambda\dagger}\, \gamma^{*}(k,\tau)\,,
\\
\hat{\sigma}^{\lambda}(\textbf{k},\tau)=b_{\textbf{k}}^{\lambda}\, \sigma(k,\tau)+b_{-\textbf{k}}^{\lambda\dagger}\, \sigma^{*}(k,\tau)\,.
\eea
The evolution equations in de Sitter(dS) space are
\begin{eqnarray}\label{eqgp}
&&\gamma^{''}-\frac{2}{\tau}\gamma^{'}+k^2 \gamma=0\, ,\\\label{eqgm}
&&\sigma^{''}-\frac{2}{\tau}\sigma^{'}+k^2 \sigma+\frac{{\tilde{\beta}}\,m^2 }{2\, \tau^2 H^2}\sigma=0\,,
\end{eqnarray}
where $'$ denotes derivatives w.r.t. conformal time, and the solutions for the mode functions read
\bea
\gamma(k,\tau)&=&\sqrt{2}\,H\frac{\left(i-k \tau\right) e^{-i k\tau}}{k^{3/2}} \,,
\label{gamma0p}
\\
\sigma(\tau,k)&=&H (1+i) \sqrt{\frac{\pi}{2}}(-\tau)^{3/2} e^{\frac{i \pi}{4}\tilde{\nu}}\,\mathcal{H}^{(1)}_{\tilde{\nu}/2}\left[-k\tau\right]\,,
\label{solgm}
\eea
with $\mathcal{H}^{(1)}_{\alpha}[x]$ the Hankel function of the first kind,
and
\bea
\tilde{\nu}\equiv\sqrt{9-\frac{2{\tilde{\beta}}\,m^2 }{H^2 }}\,. 
\eea
In deriving solutions \eqref{gamma0p}, \eqref{solgm} one assumes standard Bunch-Davies conditions for the vacuum.
We now move on to observables, starting with the power spectrum.

\subsection{Power spectrum}

\begin{figure}[ht!]
\begin{center}
  \includegraphics[width=12cm]{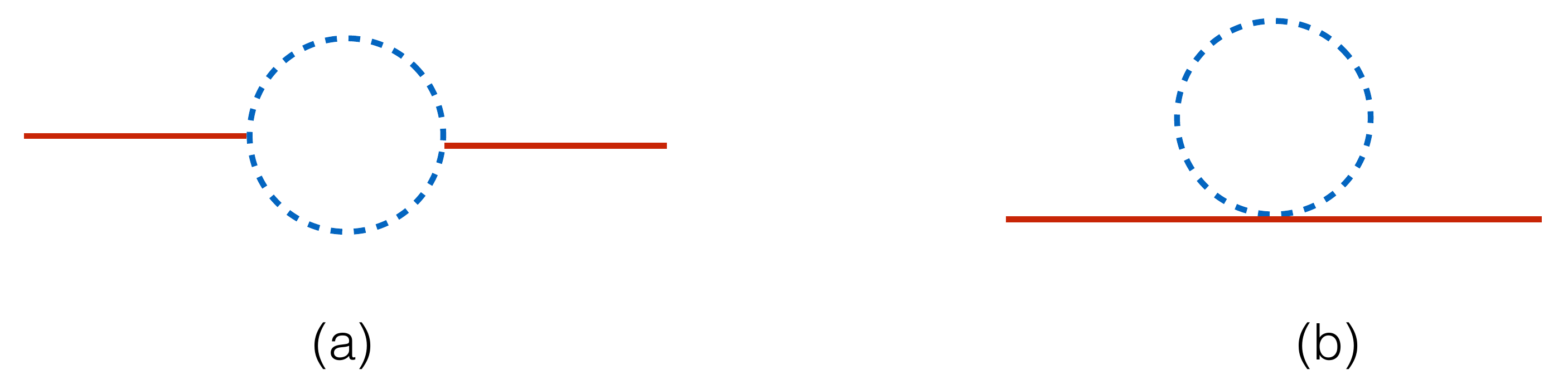}
  \caption
 {\it Diagrammatic representation of one loop contributions to the power spectrum of $\langle g^2\rangle$. Red (solid) lines indicate propagators of $\gamma$, blue (dotted) lines of $\sigma$.  }
 \label{fig2a}
\end{center}
\end{figure}

 The power spectrum of each polarization of the tensor perturbations is formally given by 
  \be
  \langle  g_{\vec k} g_{\vec k'} \rangle
  \,\equiv(2\pi)^3\,\delta^{(3)} (\vec k+\vec k')\frac{2\pi^2}{k^3}\mathcal{P}(k)\,.
  \ee
Note that we are after the power spectrum of the field $g_{ij}$ because it is the one metric that couples to matter. We shall, in the following calculations, express $g$ in the $(\gamma,\sigma)$ basis for convenience, but we will always have in mind $\langle g g \rangle $ and $\langle g g g \rangle $ as the end products of our analysis on the observables. Using Eq.~(\ref{gamma0p}), one finds the leading power spectrum to be
\begin{equation}
\mathcal{P}_{}^{\rm tree}= \frac{4}{M_{f}^{2}+M_{p}^{2}}\left(\frac{H}{2\pi}\right)^{2}\,.
\end{equation}
It follows that at tree level the power spectrum for $g$ has the same structure as in the minimal inflationary scenario\footnote{The field $g$ has both a massless and a massive component but only the former survives the late time limit at tree level.}. In order to be sensitive to the new interactions, and in the absence of extra non-derivative purely-massless terms, we need the observable at hand to probe massive fields and that necessarily requires we move to loop order.
The possible one loop contributions to the power spectrum are represented in  Fig.~(\ref{fig2a}).
 
Let us analyze the origin of the diagrams in Fig.~(\ref{fig2a}). Although the wavefunction of the field $g$ in the standard $\langle g g  \rangle $ power spectrum  receives contribution from both $\gamma,\sigma$ fields, only the former (massless) field survives in the late-time  limit. Inspection of Eq.~(\ref{cubicL}) reveals that two vertices corresponding to the same $\gamma \sigma^2$ interaction contribute to diagram \textit{(a)}. Similarly, quartic interactions of the type $\gamma^2 \sigma^2$ and $(\gamma \sigma)^2$ in Eq.~(\ref{qL}) correspond to the vertex in diagram \textit{(b)}. {As clear from the cubic and quartic Lagrangians, each interaction relevant to our set-up counts one power of the `coupling constant'  $\tilde{\beta}$. It follows that at one loop the power spectrum is limited to be at most proportional to  $\tilde{\beta}^2$ and, as we shall see, the bispectrum to be at most proportional to $\tilde{\beta}^3$. Since at a given perturbative order the maximum number of bispectrum vertices exceeds those in the two-point function, it is always possible to ensure that the three-point function  is \textit{parametrically} different from the power spectrum: this will play a key role in our model departure from standard consistency relations, see {\it Section \ref{tcr}}}.
 We estimate below the size of the 1-loop contributions to the tensor power spectrum:
 \begin{eqnarray}
&&\mathcal{P}^{\rm (a) one-loop}\propto \frac{\left(m^2M^2{\tilde \beta}\right)^2}{M_{p}^{4} M_{f}^{4}}\,=\,\mathcal{P}^{\rm tree} \Bigg[\frac{m^2 \, {\tilde \beta}}{M_f^2+ M_p^2}\Bigg]\left( {\tilde \beta} \frac{m^2}{H^2}\right),\\\label{added1}
&&\mathcal{P}^{\rm (b1) one-loop}\propto \frac{m^2M^2{\tilde \beta} H^2}{\left(M_{p}^{2}+M_{f}^{2}\right)^3}\,=\,
\mathcal{P}^{\rm tree} \frac{m^2 M_p^2 M_f^2\, {\tilde \beta}}{\left(M_{p}^{2}+M_{f}^{2}\right)^3}\,,
\\\label{added2}&&\mathcal{P}^{\rm (b2) one-loop}\propto \frac{m^2M^2{\tilde \beta} H^2}{\left(M_{p}^{2}+M_{f}^{2}\right)M_{f}^{2}M_{p}^{2}}\,
=\,\mathcal{P}^{\rm tree} \frac{m^2\,\tilde{\beta}}{M_{\text{max}}^2}
\,,
\end{eqnarray}
where $M_{\text{max}}\equiv\text{max}\{M_{f},M_{g}\}$, and we make use of the definition in Eq.~\eqref{defM}.\\ 
In the previous expressions we wrote, as representative examples,  the contribution corresponding to the $\mathcal{C}_{9}$ term of the quartic Lagrangian (see Appendix~A) in Eq.~(\ref{added1}), and the contribution corresponding to terms such as $\mathcal{C}_{6}$ and $\mathcal{C}_{7}$ in Eq.~(\ref{added2}). The term in $\mathcal{P}^{\rm (b2) one-loop}$ has been computed in the limit\footnote{Notice that instead $\mathcal{P}^{\rm (b2)}\sim \mathcal{P}^{\rm (b1)}$ for $M_f\sim M_g$ .} in which either $M_{f}\gg M_{g}$ or $M_{g}\gg M_{f}$.
Upon recalling that, as a consequence of Eq.~(\ref{finin}), the combination $m^2 \tilde \beta $ is in the neighborhood of $H^2$, the three expressions  above can be interpreted as the results of a tree level contribution times a typical $\sim H^2/M_{p,f}^2$ loop suppression\footnote{Phrased differently, this corresponds to $(P_{\rm tree}-P_{\rm 1-loop})/P_{\rm tree}\sim 1-\mathcal{O}(\tilde{\beta})\times H^2/M_{p,f}^2 $ .}. \\  
\indent Before moving on to the bispectrum calculation, we pause here to comment on the scale dependence of the power spectrum. In General Relativity (GR) an exact scale-invariance is a consequence of de Sitter symmetries. Since our set-up has a richer dynamics than GR, and given that massive tensors are known to be
characterised by  a blue spectrum, one might wonder what to expect for the index $n_{\rm T}$. The extra scale-dependence of a massive external field occurs by means of an extra (w.r.t. to their massless counterpart) dependence on powers of $k/(a H)$ at late times. However, in our case as well as e.g. quasi-single field inflation (in the scalar sector), the massive fields are not external but rather part of internal legs of Feynman diagrams, and their $k/(a H)$ behaviour is   integrated over with the leading contribution typically coming from the domain corresponding to the horizon region. As a result, one does not expect an additional scaling from internal massive lines. See also \cite{Weinberg:2006ac} for a discussion on this point.\\

\subsection{Bispectrum}
\begin{figure}[ht!]
\begin{center}
  \includegraphics[width=12cm]{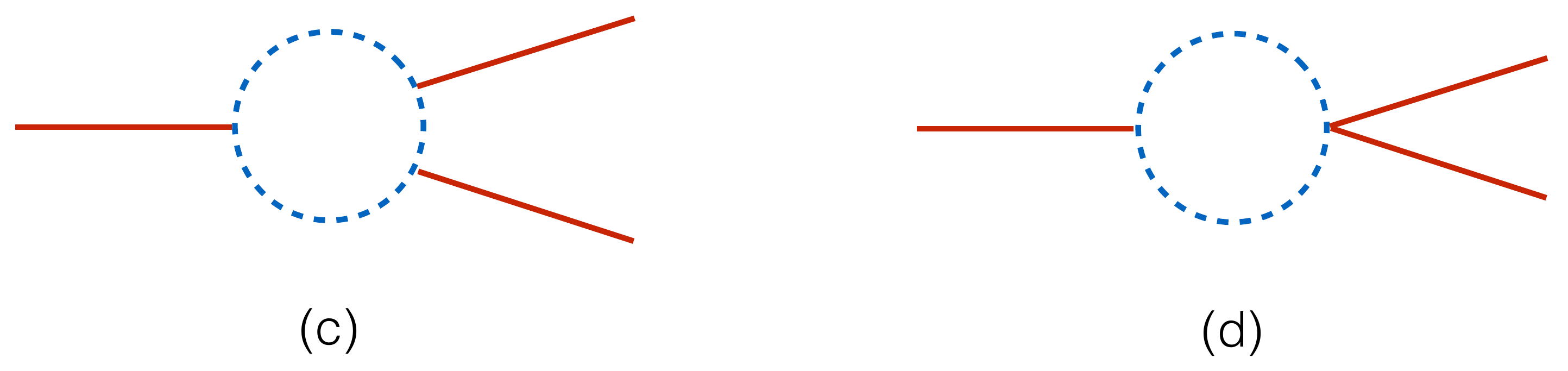}
  \caption
 {\it Diagrammatic representation of one loop contribution to the bispectrum $\langle g^3\rangle$. Red (solid) lines indicate propagators of $\gamma$, blue (dotted) lines of $\sigma$.}
 \label{fig1a}
\end{center}
\end{figure}

The main ingredients for the calculation of the 1-loop three-point function are essentially the same as for the power spectrum. The bispectrum $\mathcal{B}$ is defined as
 \be
 \langle 
   g_{\vec k_1} \,g_{\vec k_2}\,g_{\vec k_3} \rangle
  \,=(2\pi)^3 \delta^{(3)} (\vec k_1+\vec k_2+\vec k_3) {\cal B}(k_1,k_2,k_3)\,.
 \ee 
 Here too, in order to probe the extra dynamics and in the absence of new purely-massless interactions in the cubic (quartic) Lagrangian, the presence of massive spin-2 fields forces us to consider the one-loop order. Let us begin by evaluating the amplitude corresponding to diagrams (c) and (d) in Fig.~(\ref{fig1a}).
 
\noi By employing the in-in formalism,  we schematically find (see \textit{Appendix}~\ref{appC} for details):
\begin{eqnarray}\label{gena}
{\cal B}^{(c)}(k_1,k_2,k_3)&\propto& \gamma(k_{1},0)\gamma(k_{2},0)\gamma(k_{3},0)\left(m^2M^2{\tilde \beta}\right)^3\nonumber\\&&\times \int d^3q \int d\tau_{1}\,a^{4}(\tau_{1})\gamma(k_{1},\tau_{1}) \sigma(q,\tau_{1})\sigma(|\textbf{k}_{1}-\textbf{q}|,\tau_{1})\nonumber\\&&\times\int d\tau_{2}\,a^{4}(\tau_{2})\gamma(k_{2},\tau_{2}) \sigma(q,\tau_{2})\sigma(|\textbf{k}_{2}+\textbf{q}|,\tau_{2})\nonumber\\&&\times\int d\tau_{3}\,a^{4}(\tau_{3})\gamma(k_{3},\tau_{3}) \sigma(|\textbf{k}_{2}+\textbf{q}|,\tau_{3})\sigma(|\textbf{k}_{1}-\textbf{q}|,\tau_{3})
\,,\\\label{genb}
{\cal B}^{(d)}(k_1,k_2,k_3)&\propto& \gamma(k_{1},0)\gamma(k_{2},0)\gamma(k_{3},0)\left(m^2M^2{\tilde \beta}\right)^2\nonumber\\&& \times \int d^3q \int d\tau_{1}\,a^{4}(\tau_{1})\gamma(k_{1},\tau_{1}) \sigma(q,\tau_{1})\sigma(|\textbf{k}_{1}-\textbf{q}|,\tau_{1})\nonumber\\&&\times\int d\tau_{2}\,a^{4}(\tau_{2})\gamma(k_{2},\tau_{2})\gamma(k_{3},\tau_{2}) \sigma(q,\tau_{2})\sigma(|\textbf{k}_{1}-\textbf{q}|,\tau_{2})\,.
\end{eqnarray}

\noi The overall amplitudes of the bispectrum contributions are parameterized as:
\begin{eqnarray}\label{ampa}
{\cal B}^{(c)}&\propto&\frac{\left(m^2M^2{\tilde \beta}\right)^3}{M_{p}^{6} M_{f}^{6}}\,{f_{\mathcal{B}_c}(\tilde{\nu})}\,,\\\label{ampa}
{\cal B}^{(d1)}&\propto&\frac{\left(m^2M^2{\tilde \beta}\right)^2H^2}{\left(M_{p}^{2}+M_{f}^{2}\right)^3 M_{p}^{2} M_{f}^{2}}\,{f_{\mathcal{B}_{d1}}(\tilde{\nu})}\,,\\
{\cal B}^{(d2)}&\propto&\frac{m^4\,H^2\,\tilde{\beta}^{2}}{M_{\text{max}}^{6}} {f_{\mathcal{B}_{d2}}(\tilde{\nu})}\,  \,,
\end{eqnarray}
where again, diagram $(d1)$ arises from the contribution proportional to $\mathcal{C}_{9}$ in the quartic Lagrangian and diagram $(d2)$ is due to terms such as $\mathcal{C}_{6}$ and $\mathcal{C}_{7}$, with $M_{\text{max}}\equiv\text{max}\{M_{f},M_{g}\}$.
\noi Notice that, although we are able to extract almost all the dependence on the theory parameters from the integrals, there remains an irreducible dependence $f_{\mathcal{B}}(\tilde{\nu})$  which cannot be factored out but follows from numerical integration. The value of the $|f_{\mathcal{B}}(\tilde{\nu})|$ contribution it typically $\mathcal{O}(10)$ \cite{Chen:2009zp,Dimastrogiovanni:2015pla} and, naturally, a similar $f_{\mathcal{P}}(\tilde{\nu})$ functional dependence  is present also in the 1-loop power spectrum computation. The fact that $f_{\mathcal{B},\mathcal{P}}(\tilde{\nu})$, as well as their ratio, are  functions of $\tilde{\nu}$ and not directly of $\tilde{\beta}$ will be particularly important in \textit{Section} \ref{tcr}. \\
\indent We should also address the momentum dependence of the bispectrum. Standard tree level contributions of massless scalar (tensor) fields can interpolate between the local configuration ($\mathcal{B}\sim 1/(k_1^3 k_2^3)$) and the equilateral profile (with a maximum for $k_{1,2,3}\sim  1$). One typically moves from the local to the equilateral shape as the interactions considered contain an increasing number of derivatives acting on the fields \cite{Babich:2004gb,Cheung:2007st,Bartolo:2010bj,Chen:2010xka,Lewis:2011au,Fasiello:2014aqa}.  For massless fields, non-derivative interactions such as the ones we consider here, peak in the local configuration. In our set-up we are instead dealing with a contribution from massive fields running in the loop. As mentioned above, the effect of massive particles at tree level is a $(k_{\rm L}/k_{\rm S})^{3/2-\tilde{\nu}/2}$ factor in front of what would otherwise be a purely local shape \cite{Chen:2009zp,Lee:2016vti}. We therefore expect a similar mildening of the local shape in our case.
\subsubsection{Tensor non-Gaussianities}
\noindent Let us turn to providing an estimate of the various contributions to the simplest tensor non-Gaussianities, the bispectrum. We will, as ever, focus on the non-derivative contributions as we are after the effect of an extra massive spin-2 field during inflation. We estimate the size of the tensor bispectrum by means of the dimensionless parameter $f_{\rm nl}$ defined as the ratio between the three-point function and the tensor two-point correlator squared\footnote{Naturally, when it comes to the power spectrum we consider the leading tree-level contribution.}:
\begin{eqnarray}
&&f_{\rm nl}^{(c)}\propto\frac{{\cal B}^{(c)}}{\mathcal{P}^2} \propto \frac{\left(m^2M^2{\tilde \beta}\right)^3 (M_f^2+M_p^2)^2}{H^4 M_f^6 M_p^6 }  \,=\, \frac{m^6\,\tilde \beta^3}{ H^4 \left( M_f^2+M_p^2\right)}  \,,\\
&&f_{\rm nl}^{(d1)}\propto \frac{{\cal B}^{(d_1)}}{\mathcal{P}^2} \propto \frac{\left(m^2M^2{\tilde \beta}\right)^2 }{\left(M_{p}^{2}+M_{f}^{2}\right)H^2\,M_{p}^{2} M_{f}^{2}}     
 \,=\, \frac{m^4\,\tilde \beta^2\, M_{p}^{2} M_{f}^{2} }{ H^2 \left( M_f^2+M_p^2\right)^3}
\,,
\end{eqnarray}
where we have used Eq.~\eqref{defM}. 
Now, for the purposes of providing an overall consistent estimate of the bispectrum amplitude, we remind the reader of Eq  (\ref{finin}): such inequalities limit the bispectrum contributions to being respectively of order
\bea 
f_{\rm nl}^{(c)}\lesssim 10^3 \times \frac{H^2}{M^2} \; ; \qquad  f_{\rm nl}^{(d)}\lesssim 10^2 \times \frac{H^2}{M^2} \; ;
\label{fnlbounds}
\eea
 where in providing both estimates we have taken $\tilde{\beta}\sim\mathcal{O}(10)$ and used $M^2$ as a place order for the ``Planck" masses $M_{f,p}$ and combinations thereof. The significance of Eq.~(\ref{fnlbounds}) is intuitively clear: a loop-suppression is slightly lessened by the possibility\footnote{Notice that we are allowed to scan such values for the dimensionless coupling $\tilde{\beta}$ because we can trust the validity of the ghost-free theory all the way to scales where non-linearities are as important as the linear theory.} 
of an order $\mathcal{O}(10)$ value for $\tilde{\beta}$. 
As we shall see in \textit{Section} \ref{tcr}, the bispectrum and in particular its squeezed limit are an ideal probe of the presence of extra-fields during inflation. The soft three-point function carries information on the mass, the spin and the coupling of any additional inflationary particle content \cite{Jeong:2012df,Arkani-Hamed:2015bza}. As one goes up the spin ladder, bounds  such as the Higuchi inequality, essentially enforcing $m_{\rm extra}\sim H$,  are unavoidable. One would be tempted to conclude then that these extra fields will naturally be short-lived and their impact can only be an integrated-over time effect. This is precisely what we find in our set-up. On the other hand, it worth pointing out at this stage that a proven way to prolong the \textit{direct} effect of massive fields is to non-minimally couple them to massless degrees of freedom. This coupling is familiar, for example, from vector inflation models where, massive, otherwise-decaying vector modes are lifted by direct coupling to the inflaton \cite{Sorbo:2011rz}.\\
\indent We have specifically chosen in this work to focus on the simplest, non-linearly consistent theory of a (necessarily massive) extra spin-2 field, sure in the knowledge that ours is a ghost-free model. We leave the exploration of a related model with non-minimal coupling to future work \cite{future}.

%%%%%%%%%%%%%%%%%%%%%%%
\section{Tensor Consistency Relation}
\label{tcr}
%%%%%%%%%%%%%%%%%%%%%%

\subsection{The standard case}
\begin{center}
\begin{figure*}[ht]
 \includegraphics[scale=0.50]{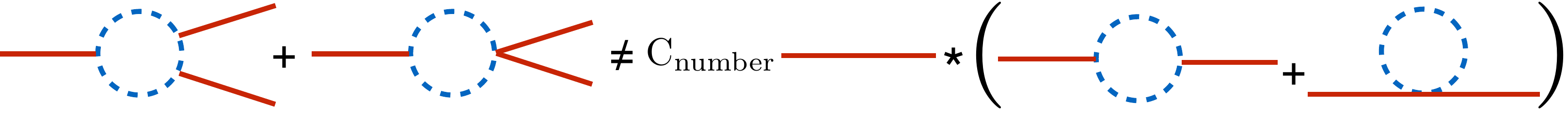}
\caption{\it Diagrammatic representation of LHS and RHS of what would be, should the equality sign hold, the consistency relation for the 
squeezed limit of the bispectrum including 
non-derivative interactions.} 
\label{ccs}
\end{figure*}
\end{center}

Before specializing to our case, let us report here the standard tensor consistency relation (CR) \cite{Maldacena:2002vr,Sreenath:2014nka}:
\bea \label{crbis1}
{\cal B}(q,\,k,\,k)_{q\rightarrow 0} \sim \left(-\frac{3}{8}+\frac{1}{8}\frac{d \ln\mathcal{P}_{\gamma}}{d \ln k} \right) P_{\gamma}(k) P_{\gamma}(q)\; ;
\eea
As we have done throughout the text, we restrict our considerations to the non-derivative contributions to the 1-loop bispectrum and 1-loop power spectrum. Consistency of perturbation theory dictates that we consider the diagrams in Fig.~(\ref{ccs}), that is we compare the 1-loop bispectrum with a product of the tree level and 1-loop two-point function.

\noi For convenience, we reproduce here the results in Eq~\eqref{ampa} of the previous section, the amplitude of the one-loop contributions to the bispectrum:  
\bea
&&\mathcal{B}^{(c)}\sim 
\left(\frac{m^2\,\tilde \beta}{M_f^2+M_p^2} \right)^3
%\frac{\left(m^2M^2{\tilde \beta}\right)^3}{M_{p}^{6} M_{f}^{6}}\,\frac{1}{k^6}\,
 f_{\mathcal{B}_{(c)}}(\tilde{\nu})
\nonumber \\
%&&\mathcal{B}^{(a)}\sim \frac{\left(m^2M^2{\tilde \beta}\right)^3}{M_{p}^{6} M_{f}^{6}}\times \Big[\tilde{\gamma}_{+} \Big]_{\rm ext}^3 \times\int_q\; \int_{\tau_1}\int_{\tau_2}\int_{\tau_3} \tilde{\gamma}^{\;\,3}_{+}\;\cdot\;\tilde{\gamma}^{\;\,6}_{-}\nonumber \\
&&\mathcal{B}^{(d1)}\sim  
\frac{H^2 \,M_f^2 \,M_p^2\,m^4\,\tilde \beta^2}{\left( M_f^2+M_p^2 \right)^5}
%\frac{\left(m^2M^2{\tilde \beta}\right)^2H^2}{\left(M_{p}^{2}+M_{f}^{2}\right)^3 M_{p}^{2} M_{f}^{2}} \,\frac{1}{k^6}\, 
f_{\mathcal{B}_{(d1)}}(\tilde{\nu})\;,
\label{expbis}
\eea
where one need not specify the  form of  the $f$ functions, but we do underscore their $\tilde{\nu}$ dependence. In order to probe the consistency relation, the squeezed limit of the bispectrum ought be compared with the power spectrum. We stress here that it is not the purely-tree-level bispectrum nor, correspondingly, the purely-tree-level power spectrum we are after: both observables are insensitive to the presence of new interactions mediated by massive modes and will therefore be connected by the standard CR. In order to capture departures from standard tensor CRs, one ought to compare the (squeezed) one-loop bispectrum to one loop contributions to the squared power spectrum: 

\bea
&&\mathcal{P}^{\rm tree} \times \mathcal{P}^{(a){\rm 1-loop}}= 
 \frac{\left(H\,m^2\,{\tilde \beta}\right)^2}{\left(M_{p}^{2}+ M_{f}^{2} \right)^3} \,
 %\frac{1}{k^6}
  \, f_{\mathcal{P}_{(a)}}(\tilde{\nu})\,,
 \nonumber \\
&&\mathcal{P}^{\rm tree} \times \mathcal{P}^{(b1){\rm 1-loop}}= \frac{H^2\,m^2\,M_f^2\,M_p^2\,\tilde \beta}{\left( M_p^2+M_f^2\right)^5}
%\times
% \frac{m^2M^2{\tilde \beta} H^2}{\left(M_{p}^{2}+M_{f}^{2}\right)^3} \,
 %\frac{1}{k^6}\,
  f_{\mathcal{P}_{(b)}}(\tilde{\nu})\, . 
  %\nonumber\\
\label{expps}
\eea 
Comparing the results of Eqs.~\eqref{expbis}-\eqref{expps}, one realizes that the consistency relation in Eq.~\eqref{crbis1} does not hold in general, since the overall dependence on the parameter $\tilde \beta$ is different in the two cases: up to $\tilde \beta^3$ for the bispectrum, but only up to $\tilde \beta^2$ for the power
spectrum. It is important to note that the functions $f_{\mathcal{B},\mathcal{P}}$ cannot make up for the ``missing" $\tilde{\beta}$ power: we know by inspection that $\tilde{\nu}$ is not a linear (or polynomial) function of simply the $\tilde{\beta}$ variable and therefore a ratio of different $f$-functions cannot restore the consistency relation.\\
\indent In our set-up then, the standard consistency relation does not hold. This should not come as a surprise in a multi-field model; after all, it is a very well-known result that a large $f^{\rm loc.}_{\rm nl}$ would be, if detected,  a smoking gun for multi(scalar)-field inflation. One might well argue that in  multi-field models the extra scalar is light, $m\ll H$ , whilst in our case the extra field is massive. There is however a scalar correspondent for precisely our case, namely quasi-single-field inflation, where for the extra field $\sigma$ one has $m_{\sigma}\sim H$. Consistency relations are indeed modified in QsF as well. We find it worthwhile to provide, alongside the ``brute force" proof of modified consistency relation just above, a general presentation on the conditions that grant CR modification/violation and place our result in a wider context. We do so in \textit{Section} \ref{gpcr1}, \ref{gpcr2} and point out similarities and subtle differences with respect to the QsF case along the way.

\subsection{General Perspective on Consistency Relations}
\label{gpcr1}
In what follows we will cast the modifications/violations of consistency relations found in our set-up in a wider context and then briefly draw a parallel with the case of CRs in QsF inflation. Consistency relations stem from a residual gauge symmetry in a physical system description\footnote{It is sufficient for symmetries to be in place at the level of the equations of motion, such as is often the case in large scale structure \cite{Kehagias:2013yd,Peloso:2013zw}.}. Let us describe (along the lines of \cite{Assassi:2012zq}) the action of one such gauge transformation ``a"  on a specific  field as the action of the charge $Q_{a}$ associated to the transformation. We have
\bea
\delta_{a} \gamma = i[Q_{a},\gamma]= \underbrace{(...)\, \gamma}_{\rm linear}+ \underbrace{(\dots)}_{\rm non-linear},
\eea
where, for the sake of generality, we do not {yet} specify the form of the transformation. It suffices here to stress that, upon acting on the field $\gamma$, the transformation may generate a linear and also a non-linear component in the field and it is the latter that will be crucial for CRs. The action of the transformation ``a" on the power spectrum built out of gamma fields will be, schematically,
\bea
\delta_a\cdot \langle \gamma \gamma \rangle \propto 
 \bcancelto{_{0}}{\langle \gamma \rangle}
  +{\rm Diff}\cdot \langle \gamma \gamma \rangle
\label{CS1}
  \eea
where the first term on the RHS vanishes\footnote{Note that for generic n-point functions the result of the non-linear transformation on the observable $\mathcal{O}_{\gamma}$ does not vanish, rather it results in non-connected diagrams \cite{Horn:2014rta} which do not play a role in the CRs we are interested in.} and we have indicated a generic differential operator (see Eq.({\ref{dilatationCS}}) for a specific example) as \textit{Diff}. The second term in Eq.~(\ref{CS1}) is the result of the linear component of the gauge transformation on $\gamma$. Let us set aside the result of the gauge transformation acting on $P_{\gamma}$ as obtained in the RHS of Eq.~(\ref{CS1}) and re-write the action of ``a" as:
\bea
\delta_a\cdot \langle \gamma \gamma \rangle = i[\langle Q_{a},\gamma\gamma \rangle ] \propto \sum_{n} \langle Q_{a}|n\rangle\langle n| \gamma\gamma \rangle ,
\label{CS2}
\eea 
where we have inserted the identity operator ${1\!\!1}= |1_{\gamma}\rangle\langle 1_{\gamma}|+  \sum_{n\geq2} |n\rangle\langle n| $. Note that we have explicitly singled out the state $|1_{\gamma} \rangle\equiv |\gamma \rangle$  to underline that the right hand side of Eq.~(\ref{CS2}) has at least one non-trivial contribution. Putting together Eqs.(\ref{CS1}, \ref{CS2}), one finds that 
\bea
{\rm Diff}\cdot \langle \gamma \gamma \rangle \propto  \langle Q_{a}|\gamma\rangle\langle \gamma| \gamma\gamma \rangle + \sum_{n\geq 2} \langle Q_{a}|n\rangle\langle n| \gamma\gamma \rangle
\label{CS3}
\eea
Switching to Fourier space, and {using the invariance of the vacuum under the linear component of} the transformation ``a", one finds that the gauge transformation obeys $\langle Q_{a}|\gamma_q\rangle\propto (2\pi)^3 \delta(\bf{q}) $, and using the overall momentum conservation:
\bea
{\rm Diff}\cdot \langle \gamma_{\bf k} \gamma_{\bf -k} \rangle \propto  c_{\rm number}\langle \gamma_{\bf{q}\rightarrow 0}| \gamma_{\bf k}\gamma_{\bf -k} \rangle + \sum_{n\geq 2} \langle Q_{a}|n\rangle\langle n_{\bf{q}\rightarrow 0}| \gamma_{\bf k}\gamma_{\bf -k} \rangle\; .
\label{CS4}
\eea
If the field $\gamma$ is the only one transforming non-linearly under $Q_a$, each term of the sum over $n\geq2$ on the right hand side vanishes and the reader will recognize in Eq.(\ref{CS4}) the standard form of the consistency relation between the squeezed (soft) limit of the three-point function (RHS) and the variation of the two-point function (LHS) of the corresponding two hard modes. Crucially, whenever (i) there exist at least one additional field\footnote{All fields $|n \rangle$ need to be suitably orthogonalized given that they have been introduced as defining the identity operator.} with a non-linear component under the symmetry, the second term in the RHS of Eq.(\ref{CS4}) can be non-zero and reveal a modified consistency relation. By inspection of Eq.(\ref{CS4}), it is clear that one must then also require that $\langle n_{\bf{q}\rightarrow 0}| \gamma_{\bf k}\gamma_{\bf -k} \rangle\not=0$. However, this is not yet satisfactory: it ought also be the case that (ii) such expression is non-trivial, in the sense that:
\bea
\langle n_{\bf{q}\rightarrow 0}| \gamma_{\bf k}\gamma_{\bf -k} \rangle\not= c_{\rm number}\times \langle n_{\bf{q}\rightarrow 0}|\gamma_{\bf{q}\rightarrow 0}\rangle \times \langle \gamma_{\bf{q}\rightarrow 0} |\gamma_{\bf k}\gamma_{\bf -k} \rangle\; ,
\label{CS5}
\eea
something that would amount to a trivial redefinition of the standard consistency relation. The immediate, observational, consequence of a modified CR is that the squeezed limit of a specific three-point function can no longer be absorbed by a gauge transformation: it is only a specific linear combination\footnote{Strictly speaking, consistency relations are typically true only at leading order in powers of $q/k$. A more precise statement is that standard CRs, when in place, make the leading contribution to the squeezed bispectrum a gauge artifact.} of at least two three-point functions that corresponds to a gauge artifact. 

The power of the perspective we have just outlined on CR is best illustrated in the set-up of quasi-single-field inflation, see \cite{Assassi:2012zq}. We shall refer the reader to existing literature on QsF, with the exception of providing here the specific example of the gauge transformation acting on the scalar field $\zeta$ in the case of QsF:
 \bea
[Q_d, \zeta]\sim -1- x\cdot \partial_{x} \zeta\; .
\label{dilatationCS}
\eea
\noi The transformation ``d" at the heart of the scalar CRs, is the dilatation symmetry (see the work in \cite{Assassi:2012zq}), and the associated charge is $Q_d$. It is straightforward to prove that the orthogonalized component of the extra massive scalar field in QsF, $\sigma^{\perp}$, has both (i) a non-linear transformation under dilatation and (ii) non-trivial (self-)interactions guaranteeing the condition in Eq.(\ref{CS5}) is satisfied. Does the same apply to our set-up?  We address this next.

\subsection{Extra massive spin-2 field during inflation}
\label{gpcr2}
The first condition we have outlined for a CR modification is the existence of two fields non-linearly transforming under the symmetry generating the consistency relation. In the case of tensors, it is a well-known result that the  gauge transformation in question is anisotropic rescaling \cite{Hinterbichler:2013dpa}. The fluctuations of tensor fields in the $(f_{\mu\nu},g_{\mu\nu})$ basis do transform also non-linearly,
\bea
\begin{split}
& \delta \gamma_{g \, \mu \nu } = a^{-2} \left(\xi^\alpha \partial_\alpha  {g}^{(0)}_{ \,
  \mu \nu} +   g^{(0)}_{ \, \alpha \nu}  \, \partial_\mu
  \xi^\alpha +  g^{(0)}_{ \, \mu \alpha}  \, \partial_\nu \xi^\alpha \right)
\, , \\
\vspace{0.1cm}
&\delta \gamma_{f \, \mu \nu } = b^{-2} \left(\xi^\alpha \partial_\alpha \, {f}^{(0)}_{ \,\mu \nu } +   f^{(0)}_{ \, \alpha \nu}  \, \partial_\mu
  \xi^\alpha +  f^{(0)}_{ \, \mu \alpha}  \, \partial_\nu \xi^\alpha \right) \, ,
\end{split}
\eea
and so do their traceless transverse components.
Note that $\xi^{\alpha}$ is the parameter of the most general gauge transformation and $a,b$ the scale factors for the metrics $g_{\mu\nu},f_{\mu\nu}$ respectively. Since we are after the CR involving the soft limit of $\langle \gamma_g \gamma_g \gamma_g\rangle $, the next step is to require that there exist $\gamma_f$-mediated interaction(s) that contribute to the bispectrum through a diagram that cannot be written as the RHS of Eq.(\ref{CS5}) upon identifying $|n\rangle\equiv |\gamma_f^{\perp}\rangle$ and $|\gamma\rangle\equiv |\gamma_g\rangle$. The existence of the cubic and quartic $\gamma_f$ self-interactions detailed in \textit{Section} \ref{hellohello} suggests that this is indeed the case.\\
\indent All ingredients seem to be in place for us to be able to declare a modification of tensor consistency relations in our set-up. There is, however, one more piece to the puzzle  with respect to the QsF case: there, a functionally independent $\sigma$ cubic self-interaction $V^{\prime \prime \prime}(\sigma)$ guarantees that the contribution on the LHS of Eq.(\ref{CS5}), when mediated by $V^{\prime \prime \prime}(\sigma)$, cannot be obtained by means of $\zeta$-$\sigma$ interactions and $\zeta$ self-interactions. In our case things are not as straightforward given that all tensor (mixed and self-)interactions share a similar dependence on the theory parameters $m^2$, $\beta_n$s. It is then necessary to resort to the more  detailed calculation presented in \textit{Section} \ref{tcr}, where we have provided a proof that the CR is indeed modified parametrically.\\
Another line of reasoning that arrives at the same conclusion goes as follows. Consistency relations during inflation stems from the theory being invariant under space diffs \cite{Berezhiani:2013ewa}. It is well-known that a massive graviton necessarily breaks such diffs and this is a sufficient condition for CRs breaking. One may always restore diff invariance by introducing extra-fields\footnote{But these are precisely the fields that will transform non-linearly!} through the Stueckelberg ``trick" \cite{ArkaniHamed:2002sp}. However, this process makes the extra degrees of freedom (due to the diff breaking) manifest and these will nevertheless eventually modify CRs.

\subsection{Quadrupolar tensor anisotropy}

\noindent An inflationary correlation between a long-wavelength ($K$) and two short-wavelength ($k_{1}\approx k_{2}\approx k$) modes induces a modulation of the power spectrum by the long-wavelength fluctuation. If the long-wavelength mode is a tensor perturbation, the induced modulation is of a quadrupolar type, as it has been shown for density fluctuations \cite{Dai:2013kra,Brahma:2013rua,Dimastrogiovanni:2014ina} (see also \cite{Giddings:2010nc}):

\begin{equation}\label{q1}
P_{\zeta}(\textbf{k},\textbf{x})=P_{\zeta}(k)\left[1+ \mathcal{Q}_{ij}(\textbf{x})\hat{k}_{i}\hat{k}_{j} \right] \,,
\end{equation}
where $P_{\zeta}(\textbf{k},\textbf{x})\equiv\int d^3\textbf{x}^{'} \,e^{-i\textbf{k}\cdot\textbf{x}^{'}}\langle\zeta(\textbf{x})\zeta(\textbf{x}+\textbf{x}^{'})\rangle_{\gamma}$ is the power spectrum for small-scale modes in a region (causal patch) centered in $\textbf{x}$, in the presence of a fixed $\gamma$. Eq.~(\ref{q1}) is valid to leading order in $(K/k)$.\footnote{These and the following results apply under the assumption that $|\textbf{x}^{'}|\ll 1/K$ (or, in other words, of a long-wavelength mode longer than the size of the local patch), and with the triangular inequality, $\textbf{K}=-\textbf{k}_{1}-\textbf{k}_{2}$, in place.} In Eq.~(\ref{q1}) we defined
\begin{equation}\label{q2}
\mathcal{Q}_{ij}(\textbf{x})\equiv\sum_{\lambda}\int_{L^{-1}}\frac{d^3 K}{(2\pi)^3}e^{i\textbf{K}\cdot\textbf{x}}\frac{\mathcal{B}_{\gamma\zeta\zeta}^{\lambda}(\textbf{K},\textbf{k},-\textbf{K}-\textbf{k})}{P_{\gamma}(K)P_{\zeta}(k)}\gamma^{\lambda}(\textbf{K})\epsilon^{\lambda}_{ij}(\textbf{K})\,,
\end{equation}
with $\lim_{K\ll k_{1}\approx k_{2}}\langle \gamma_{\textbf{K}}^{\lambda}\zeta_{\textbf{k}_{1}}\zeta_{\textbf{k}_{2}}\rangle\simeq-(2\pi)^3\delta^{(3)}(\textbf{K}+\textbf{k}_{1}+\textbf{k}_{2})\mathcal{B}^{\lambda}_{\gamma\zeta\zeta}(\textbf{K},\textbf{k}_{1},\textbf{k}_{2})\epsilon^{\lambda}_{ij}(\textbf{K})\hat{k}_{2i}\hat{k}_{3j}$ and $L^{-1}$ indicating an infrared cutoff. We note here that the bispectrum in Eq.~(\ref{q2}) is to be understood as the directly observable one, that is, the leading part of the squeezed bispectrum that cannot be removed by a gauge transformation. This is automatically the case whenever the contributions under study violate the consistency relation. \\

\noindent The result in Eq.~(\ref{q1}) readily generalizes to tensor perturbations: the local power spectrum for tensor modes in the presence of a fixed long-wavelength tensor perturbation is given by 
\begin{equation}\label{q3}
P_{\gamma}(\textbf{k},\textbf{x})=P_{\gamma}(k)\left[1+ \mathcal{Q}_{ij}(\textbf{x})\hat{k}_{i}\hat{k}_{j} \right] \,,
\end{equation}
where now
\begin{equation}\label{q4}
\mathcal{Q}_{ij}(\textbf{x})\equiv\sum_{\lambda}\int_{L^{-1}}\frac{d^3 K}{(2\pi)^3}e^{i\textbf{K}\cdot\textbf{x}}\frac{\mathcal{B}_{\gamma\gamma\gamma}^{\lambda}(\textbf{K},\textbf{k},-\textbf{K}-\textbf{k})}{P_{\gamma}(K)P_{\gamma}(k)}\gamma^{\lambda}(\textbf{K})\epsilon^{\lambda}_{ij}(\textbf{K})\,,
\end{equation}
and $\lim_{K\ll k_{1}\approx k_{2}}\langle \gamma_{\textbf{K}}^{\lambda}\gamma_{\textbf{k}_{1}}\gamma_{\textbf{k}_{2}}\rangle\simeq-(2\pi)^3\delta^{(3)}(\textbf{K}+\textbf{k}_{1}+\textbf{k}_{2})\mathcal{B}^{\lambda}_{\gamma\gamma\gamma}(\textbf{K},\textbf{k}_{1},\textbf{k}_{2})\epsilon^{\lambda}_{ij}(\textbf{K})\hat{k}_{2i}\hat{k}_{3j}$. \\

\noindent The quadrupolar modulation can be expanded in spherical harmonics as
\begin{equation}\label{q5}
P_{\gamma}(\textbf{k},\textbf{x})=P_{\gamma}(k)\left[1+ \sum_{m=-2}^{+2}\mathcal{Q}_{2m}(\textbf{x})\,Y_{2m}(\hat{k}) \right]\,,
\end{equation}
from which one derives the following expression for the quadrupole moments \cite{Dai:2013kra}
\begin{equation}\label{q6}
\mathcal{Q}_{2m}(\textbf{x})=\frac{\int d^2\hat{k}\,P_{\gamma}(\textbf{k},\textbf{x})\,Y^{*}_{2m}(\hat{k})}{\int d^2\hat{k}\,P_{\gamma}(\textbf{k},\textbf{x})\,Y^{*}_{00}(\hat{k})}\,.
\end{equation}
The root-mean-square of the quadrupole moments is what one can constrain observationally:
\begin{equation}
\overline{\mathcal{Q}^{2}}\equiv\Big\langle\sum_{m=-2}^{+2}|\mathcal{Q}_{2m}|^2\Big\rangle\,.
\end{equation}
For $\mathcal{B}_{\gamma\gamma\gamma}(K,k,|\textbf{K}-\textbf{k}|)/(P_{\gamma}(K)P_{\gamma}(k))=\text{f}_{\gamma\gamma\gamma}$, where $\text{f}_{\gamma\gamma\gamma}$ is a constant, one finds for example
\begin{equation}
\overline{\mathcal{Q}^{2}}\propto\int\frac{d^3 K}{(2\pi)^3}\text{f}_{\gamma\gamma\gamma}^{2}\,P_{\gamma}(K)\,.
\end{equation}
\noindent The magnitude of the quadrupolar anisotropy in the scalar power spectrum has been constrained to $\lesssim 10^{-2}$ from CMB \cite{CMBquadrupole}  and large scale structure \cite{Ando:2008zza} observations. Similarly, statistical anisotropies in the tensor power spectrum may be constrained with with upcoming observations of CMB polarization anisotropies and with interferometers.

\section{Conclusions}
\label{sec-con}
Inflationary dynamics takes place at energies that can be as high as $10^{14}\,{\rm GeV}$. This simple fact makes invaluable any observable carrying imprints of such an era: there is no comparable window on such high energy processes at our disposal. In this context, late-time cosmological correlation functions are the closest thing one can find to observables we are familiar with from earthbound colliders. Much as what happens for colliders, one may ask how to spot the presence of extra\footnote{Gravity as described by GR and a minimally coupled scalar field are assumed as the minimal scenario.} physics (particles) during inflation. The ideal probe of new dynamics is the squeezed (soft) limit of the bispectrum (three-point function) as it carries information on the mass, the spin and (somewhat less directly) the coupling of the extra fields. There is a vast literature on multi-field inflation, typically focused on the extra, effectively massless, scalar field \cite{Wands:2007bd}.\\
\indent  Things are no less interesting when considering vectors, tensors and higher spin fields, given that in such cases the bispectrum acquires a non-trivial angular dependence and the power spectrum itself receives anisotropic contributions. The intriguing extra angular dependence for spin$\geq 1$ fields is mitigated by that fact that, starting with tensor modes and their ``Higuchi" bound, the allowed mass range comes with a lower limit of order $H$. The immediate consequence is that these modes are not long-lived and their effect is an integrated-over-time one. A case in point is our study on adding a spin-2 particle to the minimal inflationary content.\\
\indent Whenever the extra ingredient is a tensor mode, one can take full advantage of the fact that, for interacting spin-2 fields, a ghost-free theory exists at the fully non-linear level. We have studied such model in the inflationary context, showing how tensor consistency relations are modified in this set-up and clarifying the observational consequences that ensue. The tensor bispectrum is naturally sensitive to the presence of an extra mode which we show can also induce a quadrupolar anisotropy in the power spectrum. \\
\indent A proven way to further enhance the effects of the extra (massive) degrees of freedom is to enforce non-minimal coupling to existing long-lived massless field(s).  The well-studied $\phi\,F\tilde{F}$ mixing, coupling the inflaton to gauge fields \cite{Sorbo:2011rz}, is just one such example. A natural extension of the work presented here consists in exploring the non-minimal couplings that would make the new massive modes even more impactful on cosmological observables. In doing so one has to exert special care so as to preserve the delicate ghost-free structure of multi-metric interactions \cite{Hinterbichler:2012cn}.
It is also of interest to study how the presence of massive tensor modes reflects in particular on the \textit{shape} of non-Gaussianities in scenarios with large tensor bispectra.  We leave these directions to future work \cite{future}.

\subsection*{Acknowledgments}
ED is supported in part by DOE grant DE-SC0009946. ED would like to thank the Perimeter Institute for Theoretical Physics (Canada) for hospitality and support whilst this work was in progress. MRF acknowledges support by STFC grant ST/N000668/1. GT is  partially supported by the STFC grant ST/P00055X/1. The authors are grateful to the organizers of the 
workshop ``Understanding cosmological observations" at the Centro de Ciencias de Benasque ``Pedro Pascual", where this work was initiated. MRF thanks K.~Koyama and D.~Wands for illuminating conversations.

\begin{appendix}

\section{Quartic Lagrangian expansion} \label{app-qua}

\noindent The quartic Lagrangian in the $\left(\gamma,\sigma\right)$ basis reads

\begin{eqnarray}\label{qL}
\mathcal{L}_{4}&=&-\frac{m^2 M^2\beta}{128}\Big(\mathcal{C}_{1}\,\text{Tr}\left[\sigma^4\right]+\mathcal{C}_{2}\,\text{Tr}\left[\gamma^3\sigma\right]+\mathcal{C}_{3}\,\text{Tr}\left[\gamma\sigma^3\right]+\mathcal{C}_{4}\,\text{Tr}\left[\gamma^2\right]\text{Tr}\left[\gamma\sigma\right]+\mathcal{C}_{5}\,\text{Tr}\left[\sigma^2\right]\text{Tr}\left[\gamma\sigma\right]\nonumber\\&+&\mathcal{C}_{6}\,\text{Tr}\left[\gamma^2\sigma^2\right]+\mathcal{C}_{7}\,\text{Tr}\left[\left(\gamma\sigma\right)^2\right]+\mathcal{C}_{8}\,\left(\text{Tr}\left[\gamma\sigma\right]\right)^{2}+\mathcal{C}_{9}\,\text{Tr}\left[\gamma^2\right]\text{Tr}\left[\sigma^2\right]
\Big)
\end{eqnarray}
where
\begin{eqnarray}
&&\mathcal{C}_{1}\equiv \frac{3M_{f}^{4}-2M_{f}^{2}M_{p}^{2}+3M_{p}^{4}}{M_{f}^{4}M_{p}^{4}} \,,\quad\quad\quad\quad
\mathcal{C}_{2}\equiv \frac{32\left(M_{f}^{2}-M_{p}^{2}\right)}{M_{f}M_{p}\left(M_{f}^{2}+M_{p}^{2}\right)^2}  \,,\\&&
\mathcal{C}_{3}\equiv \frac{8\left(M_{f}^{2}-M_{p}^{2}\right)^3}{M_{f}^{3}M_{p}^{3}\left(M_{f}^{2}+M_{p}^{2}\right)^2}\,,\quad\quad\quad\quad\quad
\mathcal{C}_{4}\equiv -\frac{16\left(M_{f}^{2}-M_{p}^{2}\right)}{M_{f}^{}M_{p}^{}\left(M_{f}^{2}+M_{p}^{2}\right)^2}=-\mathcal{C}_{5} \,,\nonumber\\&&\mathcal{C}_{6}\equiv \frac{2\left(11 M_{f}^{4}-10 M_{f}^{2}M_{p}^{2}+11 M_{p}^{4}\right)}{M_{f}^{2}M_{p}^{2}\left(M_{f}^{2}+M_{p}^{2}\right)^2}\,,\,\,\,\,
\mathcal{C}_{7}\equiv  \frac{2\left(M_{f}^{4}-14M_{f}^{2}M_{p}^{2}+M_{p}^{4}\right)}{M_{f}^{2}M_{p}^{2}\left(M_{f}^{2}+M_{p}^{2}\right)^2} \,,\nonumber\\&&
\mathcal{C}_{8}\equiv   -\frac{8\left(M_{f}^{2}-M_{p}^{2}\right)^2}{M_{f}^{2}M_{p}^{2}\left(M_{f}^{2}+M_{p}^{2}\right)^2}\,,\quad\quad\quad\quad\,\,\,
\mathcal{C}_{9}\equiv \frac{16}{\left(M_{f}^{2}+M_{p}^{2}\right)^2} \,.\nonumber
\end{eqnarray}

\section{Polarization tensors}
\label{appB}
The direction of wave momentum $\hat k$ in polar basis is
\be
\hat k\,=\,\left( 
\sin \theta\,\cos \phi,\,\sin \theta\,\sin \phi,
\cos \theta
\right)\,.
\ee
We introduce vectors orthogonal to $\hat k$ and to one another
\bea
\hat u= \left( \sin \phi,-\cos \phi,0\right)\,,\quad\quad\quad
\hat v=\hat k\times\hat u\,,
\eea
hence
\bea
e^{(+)}_{ab}=\frac{\hat u_a \hat u_b-\hat v_a \hat v_b}{\sqrt 2}\,,\quad\quad\quad
e^{(\times)}_{ab}=\frac{\hat u_a \hat v_b+\hat v_a \hat u_b}{\sqrt 2}\,.
\eea
We have assumed that  $e^{(\times)}_{ab}$ is even and $e^{(+)}_{ab}$ 
is odd under parity. R/L polarisations are defined as 
\bea
e^{R}=
\frac{e^{(+)}+i\, e^{(\times)}}{\sqrt 2}\,,
\quad\quad\quad
e^{L}=
\frac{e^{(+)}-i \,e^{(\times)}}{\sqrt 2}\,.
\eea

\section{Details of in in calculation}
\label{appC}

\noindent We present here the details of the diagrams calculation. The expectation value of a generic operator $\Theta$ at time $t$ is given by%
\begin{equation}\label{32}
\left\langle \Theta(t) \right\rangle=\left\langle 0\,\Big| \left[\bar{T}\,\exp\left(i\int_{t_{0}}^{t}dt^{'}H_{I}(t^{'})\right)\right] \Theta_{I}(t)\left[T\,\exp\left(-i\int_{t_{0}}^{t}dt^{''}H_{I}(t^{''})\right)\right] \Big|\,0 \right\rangle\,,
\end{equation}
where $H_{I}$ is the interaction Hamiltonian, $\bar{T}$ and $T$ are, respectively, anti-time ordering and time-ordering operators (see e.g. \cite{Weinberg:2005vy,Adshead:2009cb,Chen:2016nrs}). The expression above can be unpacked by expanding the exponentials to the desired order, depending on the number of vertices present in the diagram one needs to compute: the number of vertices in a diagram corresponds to the total number of time integrals. Let us begin with the diagram in Fig.~\ref{fig1a}c. In this case we need three vertices, so the overall order (from $\bar{T}$ and $T$ combined) of the expansion for the exponentials will be three, which is given by the sum of four contributions:
\begin{equation}
\left\langle \Theta(t) \right\rangle_{3\rm v}=(0,3)+ (3,0)+ (1,2)+(2,1)\,,
\end{equation}
where 
\begin{eqnarray}
&&(0,3)\equiv \left\langle 0\,\Big| \left[\bar{T}\left({1\!\!1}\right)\right] \Theta_{I}(t)\left[\frac{(-i)^{3}}{3!}3!\,T\left(\int_{t_{0}}^{t}dt_{1}H_{I}(t_{1})\int_{t_{0}}^{t_{1}}dt_{2}H_{I}(t_{2})\int_{t_{0}}^{t_{2}}dt_{3}H_{I}(t_{3})\right)\right] \Big|\,0 \right\rangle   \,,\nonumber\\
&&(3,0)\equiv        \left\langle 0\,\Big| \left[\frac{(i)^{3}}{3!}3!\,\bar{T}\left( \int_{t_{0}}^{t}dt_{1}H_{I}(t_{1})\int_{t_{0}}^{t_{1}}dt_{2}H_{I}(t_{2})\int_{t_{0}}^{t_{2}}dt_{3}H_{I}(t_{3})   \right)\right] \Theta_{I}(t)\left[T\left({1\!\!1}\right)\right] \Big|\,0 \right\rangle   \,,\nonumber\\
&&(1,2)\equiv      \left\langle 0\,\Big| \left[i\,\bar{T}\left( \int_{t_{0}}^{t}dt_{1}H_{I}(t_{1})   \right)\right] \Theta_{I}(t)\left[\frac{(-i)^{2}}{2!}2!\,T\left(\int_{t_{0}}^{t}dt_{1}H_{I}(t_{1})\int_{t_{0}}^{t_{1}}dt_{2}H_{I}(t_{2})\right)\right]  \Big|\,0 \right\rangle   \,,\nonumber\\
&&(2,1)\equiv        \left\langle 0\,\Big| \left[\frac{(i)^{2}}{2!}2!\,\bar{T}\left(\int_{t_{0}}^{t}dt_{1}H_{I}(t_{1})\int_{t_{0}}^{t_{1}}dt_{2}H_{I}(t_{2})\right)               \right] \Theta_{I}(t)\left[-i\,T\left( \int_{t_{0}}^{t}dt_{1}H_{I}(t_{1})   \right)\right]  \Big|\,0 \right\rangle   \,,\nonumber
\end{eqnarray}
and ${1\!\!1}$ is the unit operator. Expanding the contributions (0,3), (3,0), (1,2), (2,1), one finds: $(0,3)=(3,0)^{*}$ and $(1,2)= (2,1)^{*}$ (fields to the left in a contraction are not starred, fields to the right are starred). One then obtains 
\begin{eqnarray}
\label{33}
\left\langle \Theta(t) \right\rangle_{3\rm v}&=&-2\,\rm{Im} \left[\int_{t_{0}}^{t}dt_{1}\int_{t_{0}}^{t_{1}}dt_{2}\int_{t_{0}}^{t_{2}}dt_{3}\left\langle  \Theta_{I}(t)H_{I}(t_{1})H_{I}(t_{2})H_{I}(t_{3})\right\rangle\right]\\\label{34}
&+&2\,\rm{Im}\left[\int_{t_{0}}^{t}dt_{1}\int_{t_{0}}^{t}dt_{2}\int_{t_{0}}^{t_{2}}dt_{3}\left\langle H_{I}(t_{1}) \Theta_{I}(t)H_{I}(t_{2})H_{I}(t_{3})\right\rangle\right]\,,
\end{eqnarray}
where ``Im'' stands for the imaginary part. The equation above applies to any diagram with three vertices.\\ 

\noindent With the same procedure, the corresponding expression for a diagram with one and two vertices can be derived: these are useful for the computation of the power spectra in Fig.~\ref{fig2a} as well as for the computation of the bispectrum contribution in Fig.~\ref{fig1a}d. We write below the final results
\begin{eqnarray}\label{int11}
\left\langle \Theta(t) \right\rangle_{2\rm v}&=&-2\,\rm{Re} \left[\left\langle\Theta(t)\int_{t_{0}}^{t}dt_{1}\int_{t_{0}}^{t_{1}}dt_{2}  H_{I}(t_{1})H_{I}(t_{2})\right\rangle\right]\nonumber\\
&+&\left[\left\langle\int_{t_{0}}^{t}dt_{1}H_{I}(t_{1})\Theta_{I}(t)\int_{t_{0}}^{t}dt_{2} H_{I}(t_{2})\right\rangle\right]\,,\\\label{int33}
\left\langle \Theta(t) \right\rangle_{1\rm v}&=&2\,\rm{Im} \left[ \Theta_{I}(t)\int_{t_{0}}^{t}dt_{1}H_{I}(t_{1})\rangle\right]\,.
\end{eqnarray}

\noindent For the bispectrum, the operator is given by
\begin{equation}
\Theta={\gamma}^{s_{1}}_{\textbf{k}_{1}}(\tau\rightarrow 0){\gamma}^{s_{2}}_{\textbf{k}_{2}}(\tau\rightarrow 0){\gamma}^{s_{3}}_{\textbf{k}_{3}}(\tau\rightarrow 0)\,,
\end{equation}
where $s_{i}$ are the polarization indices of the three gravitons.\\

\noindent The contribution to the interaction Hamiltonian determining the diagram of Fig.~\ref{fig1a}c is
\begin{equation}
H_{I}(\tau)\supset\alpha\int d^3x\, a^4(\tau)\gamma_{ij}(\textbf{x},\tau)\sigma_{jl}(\textbf{x},\tau)\sigma_{li}(\textbf{x},\tau)\,,
\end{equation}
where $\alpha\sim\,m^2  {\tilde{\beta}} $. The expansion of the fields in Fourier space reads
\begin{equation}
\gamma_{ij}(\textbf{x},\tau)=\int\frac{d^3k}{(2\pi)^3} e^{-i\textbf{k}\cdot\textbf{x}}{\gamma}_{ij,\textbf{k}}(\tau)\,,
\end{equation}
with 
\begin{equation}
{\gamma}_{ij,\textbf{k}}(\tau)=\sum_{\lambda}\epsilon_{ij}^{\lambda}(\hat{k}){\gamma}_{\textbf{k}}(\tau)\equiv\sum_{\lambda}\epsilon_{ij}^{\lambda}(\hat{k})\left[a_{\textbf{k}}^{\lambda}\, \gamma_{k}(\tau)+a_{-\textbf{k}}^{\lambda\,\dagger}\, \gamma_{k}^{*}(\tau)\right]\,,
\end{equation}
and commutation relations $\left[a^{\lambda_{1}}_{\textbf{k}_{1}},a^{\lambda_{2}\,\dagger}_{-\textbf{k}_{2}}\right]=(2\pi)^3\delta_{\lambda_{1}\lambda_{2}}\delta^{(3)}(\textbf{k}_{1}+\textbf{k}_{2})$. Similar relations hold for the $\sigma$ field. The quantity inside $\langle\rangle$ brackets on the right-hand-side of Eq.~(\ref{33}) becomes (for one of the permutations)
\begin{eqnarray}\label{56}
&&\alpha^3 {\gamma}_{\textbf{k}_{1}}^{s_{1}}(0)\,{\gamma}_{\textbf{k}_{2}}^{s_{2}}(0)\,{\gamma}_{\textbf{k}_{3}}^{s_{3}}(0)\nonumber\\&\quad\quad&\times\int^{0}d\tau_{1}\,d^{3}x_{1}\,a^{4}(\tau_{1})\int\frac{d^3 w_{1}}{(2\pi)^3}\frac{d^3 q_{1}}{(2\pi)^3}\frac{d^3 p_{1}}{(2\pi)^3}e^{-i\textbf{x}_{1}\cdot\left(\textbf{w}_{1}+\textbf{q}_{1}+\textbf{p}_{1}\right)}{\gamma}_{i_{1}j_{1}\,\textbf{w}_{1}}(\tau_{1})\,{\sigma}_{j_{1}l_{1}\,\textbf{q}_{1}}(\tau_{1})\,{\sigma}_{l_{1}i_{1}\,\textbf{p}_{1}}(\tau_{1})\nonumber\\&\quad\quad&\times\int^{\tau_{1}}d\tau_{2}\,d^{3}x_{2}\,a^{4}(\tau_{2})\int\frac{d^3 w_{2}}{(2\pi)^3}\frac{d^3 q_{2}}{(2\pi)^3}\frac{d^3 p_{2}}{(2\pi)^3}e^{-i\textbf{x}_{2}\cdot\left(\textbf{w}_{2}+\textbf{q}_{2}+\textbf{p}_{2}\right)}{\gamma}_{i_{2}j_{2}\,\textbf{w}_{2}}(\tau_{2})\,{\sigma}_{j_{2}l_{2}\,\textbf{q}_{2}}(\tau_{2})\,{\sigma}_{l_{2}i_{2}\,\textbf{p}_{2}}(\tau_{2})
\nonumber\\&\quad\quad&\times\int^{\tau_{2}}d\tau_{3}\,d^{3}x_{3}\,a^{4}(\tau_{3})\int\frac{d^3 w_{3}}{(2\pi)^3}\frac{d^3 q_{3}}{(2\pi)^3}\frac{d^3 p_{3}}{(2\pi)^3}e^{-i\textbf{x}_{3}\cdot\left(\textbf{w}_{3}+\textbf{q}_{3}+\textbf{p}_{3}\right)}{\gamma}_{i_{3}j_{3}\,\textbf{w}_{3}}(\tau_{3})\,{\sigma}_{j_{3}l_{3}\,\textbf{q}_{3}}(\tau_{3})\,{\sigma}_{l_{3}i_{3}\,\textbf{p}_{3}}(\tau_{3})\,.\nonumber\\
\end{eqnarray}
The expectation value of (\ref{56}) gives 
\begin{eqnarray}
&&\delta^{(3)}(\textbf{k}_{1}+\textbf{k}_{2}+\textbf{k}_{3})\alpha^3\gamma_{k_{1}}(0)\gamma_{k_{2}}(0)\gamma_{k_{3}}(0)\epsilon^{s_{1}}_{i_{1}j_{1}}(-\hat{k}_{1})\epsilon^{s_{2}}_{i_{2}j_{2}}(-\hat{k}_{2})\epsilon^{s_{3}}_{i_{3}j_{3}}(-\hat{k}_{3})\int d^3q_{1}\nonumber\\&\quad&\quad\times\int^{0}d\tau_{1}\,a^4(\tau_{1})\int^{\tau_{1}}d\tau_{2}\,a^4(\tau_{2})\int^{\tau_{2}}d\tau_{3}\,a^4(\tau_{3})\sum_{\lambda_{1},\lambda_{2},\lambda_{3}}\epsilon^{\lambda_{1}}_{j_{1}l_{1}}(\hat{q}_{1})\epsilon^{\lambda_{1}}_{j_{2}l_{2}}(-\hat{q}_{1})\epsilon^{\lambda_{2}}_{l_{1}i_{1}}(\hat{p}_{1})\epsilon^{\lambda_{2}}_{l_{3}i_{3}}(-\hat{p}_{1})
\nonumber\\&\quad&\quad\times
\epsilon^{\lambda_{3}}_{l_{2}i_{2}}(\hat{p}_{2})\epsilon^{\lambda_{3}}_{j_{3}l_{3}}(-\hat{p}_{2})
\gamma^{*}_{k_{1}}(\tau_{1})\sigma^{}_{q_{1}}(\tau_{1})\sigma^{}_{p_{1}}(\tau_{1})
\gamma^{*}_{k_{2}}(\tau_{2})\sigma^{*}_{q_{1}}(\tau_{2})\sigma^{}_{p_{2}}(\tau_{2})
\gamma^{*}_{k_{3}}(\tau_{3})\sigma^{*}_{p_{1}}(\tau_{3})\sigma^{*}_{p_{2}}(\tau_{3})\nonumber\\&\quad&\nonumber\\&\quad&\quad+\,47\,\,\rm permutations\,,
\end{eqnarray}
where $p_{1}\equiv|\textbf{p}_{1}|=|\textbf{k}_{1}-\textbf{q}_{1}|$ and $p_{2}\equiv|\textbf{p}_{2}|=|\textbf{k}_{2}+\textbf{q}_{1}|$. The structure of the integrals is similar to the one reported in Eq.~(\ref{gena}). \\

\noindent The result for the contribution in Eq.~(\ref{34}) can be derived in the same manner 
\begin{eqnarray}\label{64}
&&\delta^{(3)}(\textbf{k}_{1}+\textbf{k}_{2}+\textbf{k}_{3})\alpha^3\gamma_{k_{1}}^{*}(0)\gamma_{k_{2}}(0)\gamma_{k_{3}}(0)\epsilon^{s_{1}}_{i_{1}j_{1}}(-\hat{k}_{1})\epsilon^{s_{2}}_{i_{2}j_{2}}(-\hat{k}_{2})\epsilon^{s_{3}}_{i_{3}j_{3}}(-\hat{k}_{3})\int d^3q_{1}\nonumber\\&\quad&\quad\times\int^{0}d\tau_{1}\,a^4(\tau_{1})\int^{0}d\tau_{2}\,a^4(\tau_{2})\int^{\tau_{2}}d\tau_{3}\,a^4(\tau_{3})\sum_{\lambda_{1},\lambda_{2},\lambda_{3}}\epsilon^{\lambda_{1}}_{j_{1}l_{1}}(\hat{q}_{1})\epsilon^{\lambda_{1}}_{j_{2}l_{2}}(-\hat{q}_{1})\epsilon^{\lambda_{2}}_{l_{1}i_{1}}(\hat{p}_{1})\epsilon^{\lambda_{2}}_{l_{3}i_{3}}(-\hat{p}_{1})
\nonumber\\&\quad&\quad\times
\epsilon^{\lambda_{3}}_{l_{2}i_{2}}(\hat{p}_{2})\epsilon^{\lambda_{3}}_{j_{3}l_{3}}(-\hat{p}_{2})
\gamma^{}_{k_{1}}(\tau_{1})\sigma^{}_{q_{1}}(\tau_{1})\sigma^{}_{p_{1}}(\tau_{1})
\gamma^{*}_{k_{2}}(\tau_{2})\sigma^{*}_{q_{1}}(\tau_{2})\sigma^{}_{p_{2}}(\tau_{2})
\gamma^{*}_{k_{3}}(\tau_{3})\sigma^{*}_{p_{1}}(\tau_{3})\sigma^{*}_{
p_{2}}(\tau_{3})\nonumber\\&\quad&\nonumber\\&\quad&\quad+\,47\,\,\rm permutations\,.
\end{eqnarray}

\end{appendix}

\end{document}